\documentclass[aps,preprint]{revtex4}

\usepackage{amsmath,amssymb}
\usepackage[pdftex]{graphicx}

\newcommand{\ddd}{\cdot\cdot\cdot}

\def\be{\begin{equation}}
\def\ee{\end{equation}}
\def\bem{\begin{displaymath}}
\def\bea{\begin{eqnarray}}
\def\eea{\end{eqnarray}}
\def\eem{\end{displaymath}}

\newcommand{\C}{{\mathcal C}}

\def\bsp{\be\begin{split}}

\def\ra{\rangle}

\def\CN{{\cal N}}

\def\G{\Gamma}

\def\a{\alpha}

\def\g{\gamma}

\def\m{\mu}
\def\n{\nu}
\def\s{\sigma}
\def\r{\rho}
\def\l{\lambda}

\def\vp{\varphi}

\def\bR {\mathbb{R}}
\def\bZ {\mathbb{Z}}

\def\ep{\epsilon}
\def\del{\partial}
\def\half{\frac{1}{2}}
\def\id{1\!\!1}

\newcommand{\Tr}{{\rm Tr\,}}

\newcommand{\cN}{{\mathcal N}}
\newcommand{\cP}{{\mathcal P}}

\begin{document}
\title{ \vspace*{-1.8cm}
\begin{flushright}
\normalsize{CERN-PH-TH/2009-159}\\ 
\end{flushright}
\vspace{1cm}
\bf{\hspace{-1cm} Brane Embeddings in $AdS_4 \times \C\cP^3$\\~~\\~~\\ }}

\author{B.~Chandrasekhar$^{1}$\footnote{bcsekhar@fma.if.usp.br} and Binata Panda $^{2}$\footnote{ binata@iopb.res.in}~~\\~~\\~~\\}
\affiliation{$^1$Instituto de F\'{i}sica, Universidade de S\~{a}o Paulo,
C. Postal 05315-970, S\~{a}o Paulo, Brasil \\~~\\
$^2$ Institute of Physics, Bhubaneswar 751 005, India \\~~\\~~\\~~\\~~\\~~\\}



\begin{abstract}

We construct $D$-brane embeddings in $AdS_4 \times \C\cP^3$ by studying the consistency
conditions following from the pull back of target space equations of motion.
We explicitly discuss the supersymmetry preserved by these embeddings by analyzing the 
compatibility of kappa symmetry projections with the target space Killing spinors in
each case. The embeddings correspond to AdS/dCFT dualities 
involving ABJM theories with a defect. We also comment on the defect
CFT.

\end{abstract}
\maketitle

\section{Introduction}

Recently, string theory on $AdS_4\times\C\cP^3$ has enjoyed a special study, due to its appearance in a new example of the AdS/CFT duality~\cite{Maldacena:1997re,Witten:1998qj,Gubser:1998bc}. The example uncovered by Aharony, Bergman, Jafferis and Maldacena (ABJM), involves ${\cal N}=6$ superconformal ${\rm SU}(N)\times
{\rm SU}(N)$ Chern-Simons theory in three dimensions and
M-theory on ${\rm AdS}_4\times {\rm S}^7/{\mathbb Z}_k$, where $k$
is the level of the Chern-Simons action~\cite{Aharony:2008ug}. The new duality is motivated by the conjecture
that Superconformal Chern-Simons theories describe the low-energy world-volume dynamics of
multiple M2-branes~\cite{Schwarz:2004yj}. The ABJM model is characterized by two parameters -- the rank $N$
of the two gauge groups ${\rm SU}(N)$ and the integer level $k$
which is opposite for  the  gauge groups. Remarkably, there exists
an analogue of the 't Hooft limit, where $N,k\to\infty$ with the
ratio $\lambda=2\pi^2 N/k$ kept fixed. In this limit $\lambda$
becomes continuous allowing therefore for application of standard
perturbative techniques. In particular~\cite{Aharony:2008ug}, at strong coupling, i.e,
when $\lambda$ becomes large, the M-theory on ${\rm AdS}_4\times
{\rm S}^7/{\mathbb Z}_k$ can be effectively described by type IIA
superstring theory on the ${\rm AdS}_4\times \C\cP^3$ background. 
The connection between M-theory and type IIA string theory has been an exciting source
of information, especially, since the fundamental strings and D-branes of type IIA are on a unified footing in eleven dimensions. Consequently, starting from M-theory, it is possible to have an alternative description as a weakly coupled type IIA string theory, giving us the handle to bridge the non-perturbative
and perturbative features of these theories.  Low energy theory on the
M2-branes governed by non-dynamical CS-theory can have interesting relation to the dynamical Yang-Mills
living on the world-volume of $D$-branes in type IIA~\cite{Mukhi:2008ux,Ezhuthachan:2008ch}. Construction of
string and brane actions for IIA strings in $AdS_4\times \C\cP^3$ is a very interesting subject and has been
discussed extensively in~\cite{Arutyunov:2008if}-\cite{Cagnazzo:2009zh}.
An intriguing direction to explore for an extension of the new duality~\cite{Aharony:2008ug}, is to
consider CFT's with a defect, corresponding to $D$-branes in the bulk. Such theories have several 
applications in a variety of situations as discussed extensively in the context of $AdS_5 \times S^5$
~\cite{Mirabelli:1997aj}-\cite{Myers:2006qr}. One of the models  proposed in~\cite{randall} corresponds to that of a 3-brane living inside an $AdS_5$ can give rise to a warped geometry of
the form $AdS_4$. Such a brane is expected to
be a viable candidate for proposals of localized gravity~\cite{Randall:1999vf}. Interestingly, the metric
for such curved branes in AdS can be written down, allowing for a detailed analysis. \\

In this work, we study both supersymmetric and non-supersymmetric embeddings of $D$-branes in $AdS_4\times \C\cP^3$.  These embeddings can actually lead to dualities between supersymmetric AdS embeddings and conformal field theories with a defect, AdS/dCFT duality~\cite{bachas2}-\cite{johanna}. Situations leading to $AdS4/CFT_3$ with flavors have been considered 
in~\cite{Fujita:2009kw}-\cite{Fujita:2010pj} and D-branes in various $AdS_4$ models have been discussed 
in~\cite{Koerber:2007jb,Koerber:2009he}. In particular, kappa-symmetry gauge fixings of the superstring, $D0$ and $D2$-brane actions in the complete $AdS_4\times \C\cP^3$ superspace have been discussed from world-sheet perspective in~\cite{Grassi:2009yj}. Studying branes in $AdS_4\times \C\cP^3$ is also important from the point of view understanding local and non-local operators in the holographic dual theory~\cite{Chen:2008bp}.
A discussion of particle like branes was presented in~\cite{Aharony:2008ug},
both from the M-theory and IIA point of view.\\

In this article, we take the lead of~\cite{Skenderis:2002vf}, where a very general study of possible embeddings of $D$-branes in $AdS_5 \times S^5$ was undertaken.
The brane embeddings we discuss are non compact. 
This approach has the potential for a full classification of possible brane embeddings in $AdS_4\times \C\cP^3$ and also its Penrose limits. The construction and classification of branes in 
the Penrose limits, is useful for studying the quantization, 
superalgebra and spectra of open string fluctuations around both
static and time-dependent embeddings~\cite{Skenderis:2002wx,Skenderis:2002ps}. In this work, we restrict
ourselves to the construction of explicit solutions of branes in
$AdS_4\times \C\cP^3$ and leave the extension to the Penrose limit
for future (see also~\cite{AliAkbari:2010rs}). 
The most general $D$-brane field equations following from the pull over of
the equations of motion in the target space geometry were written down and explicit embeddings were 
constructed~\cite{Skenderis:2002vf}.  This type of analysis is better to understand the classification of embeddings based on the symmetry properties associated to a particular geometry, which is in this case, the
$AdS_4\times \C\cP^3$ background. The stability analysis involves finding out the number of 
target space supersymmetries compatible with the kappa symmetry projection on to the brane world
volume. This $D$-brane picture gives a nice view of the dynamics of a defect on the boundary CFT. 
It is thus interesting to consider an AdS brane in the present context, as it also lends
itself to a very natural holographic interpretation. To realize this situation, one adds additional structure on both sides of the duality: a D-brane in the bulk 
and a defect in the boundary theory. The theory on the defect captures holographically the physics of the D-brane in the bulk, and the interactions between the bulk and the D-brane
modes are encoded in the couplings between the boundary 
and the defect fields. \\

\smallskip

\noindent
The $AdS_4/CFT_3$ duality shares many common features with the well studied $AdS_5 \times S^5$-Super Yang-Mills
correspondence. In fact, using the insights from the later, many results can be obtained for the former, like for instance, the
existence of certain string solutions in $AdS_4$, as long as
the $\C\cP^3$~\cite{Drukker:2008zx} can be ignored.
In tune with the earlier AdS/dCFT dualities~\cite{KR}, in the following, we  
consider a $D4$-brane wrapping an $AdS_3 \times \C\cP^1$
submanifold of $AdS_4 \times \C\cP^3$. This configuration may be considered
as the near-horizon limit of a certain $D2-D4$ system, and the AdS/CFT 
duality is considered to act twice: both in the bulk and on the worldvolume. 
In the limit discussed in~\cite{ooguri1}, 
the bulk description can be taken to be in terms of supergravity coupled 
to a probe $D4$-brane. The dual theory is the ABJM model~\cite{Aharony:2008ug} coupled to 
a two dimensional defect. The defect theory may be associated with
the boundary of $AdS_3$ and  it should be a conformal field theory, following the
logic established for earlier brane embeddings~\cite{KR}. In fact, using the solutions
we present in this work, gives enough evidence for the existence of dCFT's in the case of
$AdS_4/CFT_3$ duality. \\

\noindent
Note : After submission of this work, we came to know of a $D2-D4$ intersection model in~\cite{Fujita:2009kw}, which has some overlap with one of our supersymmetric solutions of $D4$-brane wrapping a  $\C\cP^1$ and sitting at $\xi =0$, though our set ups do not break parity. \\

\noindent
The plan of the rest of the paper is  as follows. 
In Section-\ref{dbranembd},
we discuss the D-brane embeddings into the $AdS_4 \times \C\cP^3$ IIA background. In Section-\ref{Ap:Killing}, we use the kappa symmetry projector to determine the supersymmetry preserved by these D-brane embeddings. In Section-\ref{fieldtheory}, we discuss the field theory 
aspects of embeddings of Section-\ref{dbranembd} and
present a discussion of possible extensions for future in Section-\ref{conclusion}. In Appendix A and B, we discuss the D-brane field equations used in Section-\ref{dbranembd} and the analysis of killing spinors, respectively.\\

\smallskip

%
\section{D-brane Embeddings} \label{dbranembd}

In this section, we start by giving a heuristic discussion of brane embeddings in $ AdS_4\times \C\cP^3$
background and present some general features. The issue of stability of the embeddings is dealt with in more detail in the later sections when analyzing the supersymmetry.
The $ AdS_4\times \C\cP^3$ IIA background geometry we start with has the form~\cite{Aharony:2008ug,Nishioka:2008gz},
\be \label{adscp3}
ds_{IIA}^2=\tilde{R}^2(ds_{AdS4}^2+4ds_{CP3}^2), 
\ee 
where
\be
\tilde{R}^2=\frac{R^3}{4k}=\pi\sqrt{\frac{2N}{k}}.  
\ee
\bea \label{ads4}
ds_{AdS4}^2 & = & \frac{du^2}{u^2} + u^2 (dx \cdot dx)_3 \\
&=&\frac{du^2}{u^2} + u^2 [ -(dx^0)^2 + (dx^1)^2 + (dx^2)^2 ]
\eea

\bea \label{cp3earlier}
ds^2_{CP^3}&=&d\xi^2+\cos^2\xi\sin^2\xi\left(d\psi+\frac{\cos\theta_1}{2}d\varphi_1-
\frac{\cos\theta_2}{2}d\varphi_2\right)^2   \\ \nonumber
&& +\frac{1}{4}\cos^2\xi\left(d\theta_1^2+\sin^2\theta_1
d\varphi_1^2\right)+\frac{1}{4}\sin^2\xi(d\theta_2^2+\sin^2\theta_2
d\varphi_2^2). 
\eea
Now, we wish to introduce a $D4$-brane in this background which wraps a $\C\cP^1$ inside the 
$\C\cP^3$. This can be taken to be defined as:
\be
\xi =0
\ee
The radius of the $S^2$ which the brane wraps is ${\tilde R}$. The brane world volume lies inside
the $AdS_4$ and stretches along the direction $x^2 = x =0$. Thus, it fills the $AdS_3$ defined by
the coordinates $u,x_0,x_1$ and wrapping the $\C\cP^1$ parameterized by $\theta_1,\phi_1$.\\

\noindent
In the absence of the D4-brane,
the system has 24 unbroken supercharges, an $SO(2,1)$ Lorentz symmetry
acting on $(x_0, x_1, x_2)$ and an additional $SU(4)$
acting on $\C\cP^3$ geometrically. The isometry group of the metric (\ref{ads4}),
(\ref{cp3earlier}) preserved by the D4-brane is $SO(2,2) \times
SU(2)_V \times SU(2)_H \times U(1)$.  $SO(2,2)$ acts on $(u,x_0,x_1)$, while
$SU(2)_H$ rotates $(\theta_1, \varphi_1)$. From a field theory viewpoint $SU(2)_V
\times SU(2)_H \times U(1)$ is the unbroken R-symmetry and $SO(2,2)$ is the 2D
conformal group, suggesting that the dual field theory must be exactly
conformal. In the near horizon limit, one expects superconformal enhancement to twelve 
supercharges. We comment more on the symmetries of the defect field theory later on.\\

\noindent
Using a general ansatz for the embedding
surface as:
\be \label{x}
x = \frac{C}{u} \, \, ,
\ee
the induced metric can be seen to be $AdS_3$:
\be
ds^2  = (1 + C^2)\frac{du^2}{u^2} + u^2 [ -(dx^0)^2 + (dx^1)^2 ]
\ee
This leads to a shift in the curvature radius to $r^2 = \tilde{R}^2 (1 + C^2)$. Here, $C$ denotes
the minimum distance of the $AdS_3$ brane from the center. This distance is controlled depending on
whether some of the $D2$-branes end on a $D4$-brane and related to the ratio of their
tensions~\cite{Karch:2000gx}. Similar to the various cases
of brane embeddings considered in~\cite{Karch:2000gx}, one can check that the situation where $q$ of the $N$ $D2$-branes end on the $D4$-brane, lead to a nonzero value for $C$. This is essentially because, as a reaction to being pulled on by the $D2$-branes, the $D4$-brane position along $x$ becomes a function of $u$ and
the bending is of the form given in eqn.(\ref{x}). \\

\noindent
Notice that we are considering the case of an $M5$-brane in the AdS$_4$ $\times$ $S^7/\bZ_k$ geometry coming from $N$ $M2$ branes at a $\bZ_k$ singularity~\cite{Myers:2006qr}.  The 
case, where none of the branes are intersecting corresponding to with a $C=0$ AdS$_3$ inside the AdS$_4$
times an equatorial $S^3$ inside the $S^7$. When there is an additional world-volume flux, the bending was argued in~\cite{Karch:2000gx} to go as ${ x=C/u^2}$, describing 
the embedding inside the AdS$_4$. It was further pointed out that this behavior is only
adequate far away from the intersection and that in the $M2$ near-horizon
region the $M5$-brane bending is different. From the type IIA picture, we see from the above analysis that this is indeed true and the bending is actually as given in eqn.(\ref{x}). 
The naive analysis above was all in the probe approximation and it is important to see how the situation
changes when the back reaction effects are considered. In particular, the embedded branes are sitting at 
the top of the potential, signaling a tachyonic instability. However, the mass of the mode is still above
the Breitenlohner-Freedman bound~\cite{Breitenlohner:1982bm} and hence is stable.\\

\noindent
The DBI part of the action of the embedded $D$-brane on $S^2$, the case where $\xi = 0$ takes the form
$\sqrt{1 + q^2}$, in the units $\tilde R = 1$. Similarly, the contribution of the DBI action from $AdS_3$
and the WZ-terms can be calculated. The RR 4-form reads 
\bea
 F^{(4)} &=&\frac{3R^3}{8}\epsilon_{AdS_4} \\
&=& \frac{3R^3}{8} u^2 dx^0 \wedge dx^1 \wedge dx^2\wedge du
\eea
where $ \epsilon_{AdS_4} $ is the unit volume form of the  ${AdS_4} $  space.
The RR 2-form $F^{(2)}=d\tilde{A}$
in the type IIA string is explicitly given by
\bea \label{2form}
F^{(2)}&=&
k( -\cos\xi\sin\xi d\xi \wedge 
(2d\psi+\cos\theta_1d\varphi_1-\cos\theta_2 d\varphi_2) \\ \nonumber 
&&-\frac{1}{2}\cos^2\xi\sin\theta_1 d\theta_1\wedge  d\varphi_1 -\frac{1}{2}\sin^2\xi\sin\theta_2 d\theta_2 \wedge  d\varphi_2 )  
\eea 
Using these, we get the full result:
\be
\mathcal L \approx   -T_{D4} \; \left ( R^2 \sqrt{1 + { \pi^2
q^2} } \; u \; \sqrt{u^4 + (u')^2}  - \;  q
\; u^3 \right )
\ee
Both the cases $q = 0$ and $q \neq 0$ exist and the difference is in the AdS curvatures. One can derive the equations of motion and check the validity of solutions. We will not try to do this, as we do it in full generality below. 
Since the remaining part of spacetime is an $AdS_3$, one does not need to be
bothered by the tachyonic mode corresponding to fluctuations in $S^2$.\\

\noindent
Let us now look at the full set of equations of motion specifically for a D4-brane embedding in to an $AdS_4 \times \C\cP^3$ background.
Following the general procedure given in~\cite{Skenderis:2002vf}, for the world volume analysis of 
$Dp$-brane field equations in $AdS_n \times S^m$ space times, we discuss the embeddings of branes in
$CP^3$. A summary of the D-brane field equations is given in Appendix A.
Using (\ref{finfe}),  the $D4$-brane field equations in $AdS_4 \times \C\cP^3$ background, reduces to:
\bea 
e^{-\Phi}\partial_{i} (\sqrt{-M} \theta^{i i_1}) &=& \frac{1}{(2!)^2} 
\epsilon^{i_1 i_2 i_3 i_4 i_5 } F_{i_2 i_3}f_{ i_4 i_5 } + \frac{1}{4!} 
\epsilon^{i_1 i_2 i_3 i_4 i_5 } f_{i_2 i_3 i_4 i_5 }. \label{krfe1} 
\eea
\bea
&&\hspace{-25mm}\frac{1}{(2!)^3} 
\epsilon^{i_1 i_2 i_3 i_4 i_5 } (F_{i_1 i_2} \wedge F_{i_3 i_4})f_{i_5  m} +
\frac{1}{2! 3!} 
\epsilon^{i_1 i_2 i_3 i_4 i_5} F_{i_1 i_2} f_{i_3 i_4 i_5 m} \label{krfe2}   \\ \nonumber
&&= e^{-\Phi}\Big[- \partial_{i} (\sqrt{-M} G^{ij} \partial_{j} X^{n} g_{mn})  + 
\frac{1}{2} \sqrt{-M} (G^{ij} \partial_{i} X^{n} \partial_{j} X^{p} g_{np,m})\Big] . 
\eea
where $f_{i_3 i_4 i_5 m}$  denotes the pullback of $f$ on the first three indices, i.e. $f_{i_3 i_4 i_5 m} = \del_{i_3} X^{m_3} \del_{i_4} X^{m_4}
\del_{i_5} X^{m_5} f_{m_3 m_4 m_5 m}$ and in a similar manner for  $f_{i_5  m}$.\\

\noindent
Let us note that the solution set of these equations describes all possible
embeddings of D4-branes into the target space. It is possible to study some of the particle-like branes of~\cite{Aharony:2008ug} by considering D4-branes wrapping
a $\C\cP^2 \subset \C\cP^3$, corresponding to an M5-brane wrapping an $S^5$ in $S^7$ in M-theory. Here we describe D4-branes which wrap on $\C\cP^1$ in the $\C\cP^3$ corresponding to a defect.
Such embeddings can be realized from the following
ansatz: split the embedding coordinates $X^{m}$ into $\{ \eta^{i},
X^{\lambda} (\eta^{i}) \}$, where the worldvolume coordinates are
\be
\eta^{i} = \{ x^0, x^1,  u, \theta_1, \varphi_1 \} \label{ans1}
\ee
and the transverse scalars are 
\be
X^{\lambda} = \{ x^2(u) \equiv x(u), \xi, \psi , \theta_2, \varphi_2 \}, 
\label{ans2}
\ee
where  we relabel $x^2$ as $x$ and we assume
that the only dependence of the transverse scalars on the worldvolume
coordinates is in $x(u)$. We  switch on a worldvolume flux
\be  \label{flux}
F_{\theta_1 \varphi_1} = q \sin \theta_1.
\ee
Notice that this 2-form flux is proportional to the RR 2-form in eqn. (\ref{2form}) at the point $\xi =0$. 
With this ansatz, we proceed to  calculate all the
quantities appearing in (\ref{krfe1}) and (\ref{krfe2}); for example,
\be
\sqrt{-M} =\tilde{R}^3 u (1 + u^4 (x')^2)^{\frac{1}{2}} L_{(\theta_1 \xi)}
 \label{m1}
\ee
where prime denotes the derivative with respect to $u$ and we define,
\be
L_{(\theta_1 \xi)} =\Big(\tilde{R}^4 \cos^4\xi( \sin^2\xi \cos^2\theta_1 + \sin^2\theta_1 ) + q^2 \sin^2\theta_1  \Big)^{\frac{1}{2}}.
\label{lthetaxi}
\ee
Substituting the ansatz into (\ref{krfe2}), we find that the equations derived  from the $X^{m} = \{x^0, x^1,\psi, \theta_2, \varphi_1,\varphi_2\}$ equations are satisfied automatically by the ansatz. And the rest of the equations are discussed below.\\

\noindent
The $X^{m} = x^2 = x $ equation gives :
\bea
\partial_{u} 
\left ( \frac{L_{(\theta_1 \xi})} {(1 + u^4 (x')^2)^{\frac{1}{2}}} 
u^5 x' - q u^3 \sin \theta _1 \right ) = 0;
\eea
The $X^{m} = \xi $ equation leads to 
\bea
L^{-1}_{(\theta_1 \xi)} u (1 + u^4 (x')^2)^{\frac{1}{2}} \Big[ \cos^3\xi \sin\xi ( \cos^2\xi \cos^2\theta_1 - 2 \sin^2 \xi\cos^2\theta_1  - 2 \sin^2 \theta_1) \Big] = 0
\eea
 And the $X^{m} = \theta_1 $ equation leads to 
\bea
L^{-3}_{(\theta_1 \xi)} u (1 + u^4 (x')^2)^{\frac{1}{2}} \cos^4\xi \sin^2\xi \cos \theta_1 \sin \theta_1= 0
\eea
Moreover, the gauge field equation (\ref{krfe1}) 
leads to : 
\bea
 L^{-3}_{(\theta_1 \xi)} u (1 + u^4 (x')^2)^{\frac{1}{2}} \cos^4\xi \sin^2\xi \cos \theta_1 = 0
\eea
The  equation deriving from $u$ follows from the $x$-equation,
and that from $\theta_1 $ follows from the gauge field equation.
So, for the above ansatz, we find that the
only independent equations are the ones deriving from 
the $X^{m} = \{x, \xi\}$ equations along with the gauge field equation
(This is expected as worldvolume diffeomorphisms can be used
to eliminate $p+1 = 5$ equations. In addition the metric does not depend upon 3 coordinates $ \psi , \varphi_1, \varphi_2$.). \\

\noindent
Let us summarize the independent equations  
\bea
&&\partial_{u} 
\left ( \frac{L_{(\theta_1 \xi})} {(1 + u^4 (x')^2)^{\frac{1}{2}}} 
u^5 x' - q u^3 \sin \theta _1 \right ) = 0; \label{set1} \\
&&L^{-1}_{(\theta_1 \xi)} u (1 + u^4 (x')^2)^{\frac{1}{2}} \Big[ \cos^3\xi \sin\xi ( \cos^2\xi \cos^2\theta_1 - 2 \sin^2 \xi\cos^2\theta_1  - 2 \sin^2 \theta_1) \Big] = 0 ;\nonumber \\
\label{set2}
&& L^{-3}_{(\theta_1 \xi)} u (1 + u^4 (x')^2)^{\frac{1}{2}} \cos^4\xi \sin^2\xi \cos \theta_1 = 0
\label{gaugefield}
\eea
To solve (\ref{set1})-(\ref{gaugefield}) simultaneously, we first note that the $\xi$-equation (\ref{set2}) and the gauge field equation (\ref{gaugefield}) together can be solved  either when (i) $\xi= 0 $, which we will refer to as a maximal, as the radius of the $CP_1$ over which the brane is wrapped $\sim$  $\cos^2\xi$, or when (ii) $\xi= \frac{\pi}{2} $, which we will refer to as a minimal.
The $x$-equation  (\ref{set1}) yields
\be
x' = \frac{(qu^3 \sin \theta_1 -c)}{u^2\sqrt{ u^6 L_{(\theta_1 \xi)}^2 
- (qu^3\sin \theta _1 -c)^2}}, \label{xpm}
\ee
where $c$ is an integration constant. There are two interesting cases.\\

\begin{flushleft}
{Branes wrapping CP: {\em Case A}}
\end{flushleft}

\noindent
Let us now substitute solutions of the $\xi$-equation and gauge field equation
into (\ref{xpm}). First we focus on the case where the brane wrappings are maximal
 i.e., the case when $\xi= 0 $.
In this case  (\ref{lthetaxi}) simplifies to,
\be
L_{(\theta_1 \xi)} =(\tilde{R}^4 + q^2)^{\frac{1}{2}} \sin \theta_1
\label{lthetaxi-max}
\ee
Using (\ref{lthetaxi-max}), (\ref{xpm}) reduces to 
\be
x' = \frac{(qu^3-c)}{u^2(\tilde{R}^4 u^6 + 2cqu^3-c^2)^{\frac{1}{2}} }, \label{xppm}
\ee
and the induced metric on the brane is
\be
ds^2 = \tilde{R}^2 \Big[ u^2 (dx \cdot dx)_2 + \frac{u^4 (1+q^2)}{(\tilde{R}^4 u^6 + 2cqu^3-c^2) }du^2 + (d\theta_1^2 + \sin^2 \theta_1 d\varphi_1^2) \Big],
\label{indmetric1}
\ee
Setting $\tilde{R}= 1$, the induced metric takes a form,
\be
ds^2 =  u^2 (dx \cdot dx)_2 + \frac{u^4 (1+q^2)}{(u^3 - u_{+}^3)
(u^3 + u_{-}^3) }du^2 + (d\theta_1^2 + \sin^2 \theta_1 d\varphi_1^2) ,
\label{indmetric2}
\ee
where, $(u^6 + 2cq u^3 - c^2) = (u^3 - u_{+}^3)(u^3 + u_{-}^3)$, with $u_{+}^3, u_{-}^3 \geq 0$. The explicit form for the roots are needed later on:
\bea  \label{u3r}
u^3_{+} &=& - cq + \left | c \right | \sqrt{1+q^2}; \\
u^3_{-} &=& cq + \left | c \right | \sqrt{1+q^2}. \nonumber
\eea
\\
\noindent
Now, the $AdS_3 \times S^2$ embeddings can be found as follows. When $c = 0 $, (\ref{xppm}) integrates to the simple
expression
\be
x = x_{0} - \frac{q}{u}. \label{xq0} \, ,
\ee
This solution corresponds to the Karch-Randall embedding~\cite{KR}.
In this limit ,the induced metric is,
\be \label{d4ads3s2}
ds^2 =  u^2 (dx \cdot dx)_2 + \frac{ (1+q^2)}{u^2 }du^2 + (d\theta_1^2 + \sin^2 \theta_1 d\varphi_1^2) ,
\ee
The embedded geometry is $AdS_3 \times S^2$ when $c=0$. Also note that the embeddings exist even in the zero  flux ($q=0$) limit. The zero flux embedding must satisfy the 
zero extrinsic curvature trace condition (\ref{kcon}) 
since for this solution $J_{m}$ vanishes.\\
\\

\noindent
An explicit calculation of  the components of the second fundamental form given in (\ref{secondf}) leads to the following nonzero $ {\cal K}^{m}_{ij}$ components while all other components vanish.
\bea
&&{\cal K}^{x^2}_{x^0x^0} = x'u^3 \, , \, {\cal K}^{x^2}_{x^1x^1} = -x'u^3\, , \,
{\cal K}^{x^2}_{uu} = \frac{-3x'}{u} - x'' \, , \, {\cal K}^{u}_{uu} = x'^2u^3 \\ \nonumber
&&{\cal K}^{\xi}_{\theta_1 \theta_1} = \frac{-1}{4} \cos\xi \sin\xi \, , \,
{\cal K}^{\xi}_{\theta_1 \varphi_1} = \frac{1}{4} \cos\xi \sin\xi [  \cos^2 \theta_1 \cos 2 \xi - \sin^2 \theta_1] \\ \nonumber
&&{\cal K}^{\psi}_{\theta_1 \varphi_1} = \frac{-1}{4} \cos^2 \xi \sin \theta_1 - \frac{\sin^2 \xi}{2 \sin \theta_1}- \frac{\cos^2 \theta_1}{\sin \theta_1} \cos^2 \xi \\ \nonumber
&& {\cal K}^{\varphi_1}_{\theta_1 \varphi_1} = \cot \theta_1 - \frac{1}{4}[3 + \cos 2\xi]  \cot \theta_1 \, , \,
{\cal K}^{\theta_1}_{ \varphi_1\varphi_1} = \cos \theta_1 \sin \theta_1 [\cos^2 \xi -1]
\eea
For consistency,
one can check that the trace of the second fundamental form of the embedding vanishes.
Note that when $q=0$ from (\ref{xq0}), $x$ being a constant, those $ {\cal K}^{m}_{ij}$ components vanish which are functions of $x'$ or $x''$. Again, for the specific embedding using $\xi = 0$, rest of the above listed components vanish except ${\cal K}^{\psi}_{\theta_1 \varphi_1}$. Thus, unlike the similar embeddings considered in~\cite{Skenderis:2002vf}, the ones
corresponding to $D4$-branes wrapping an $S^2$ in the present context are not totally
geodesic~\cite{Bachas:1999um}. This is due to the marked difference in the sphere and $\C\cP^3$ geometries. The second fundamental form is an important quantity when studying the possible corrections
to the DBI action at higher orders in $\alpha'$. 
Since, the second fundamental form does not vanish, it modifies the pull back of the ambient curvature
tensor and contributes to processes involving the scattering of closed and open strings.
\\

\noindent
We now see how the asymptotically $AdS_3 \times S^2$ embeddings look like.
For the general case when $c \neq 0$, the induced geometry is asymptotically $AdS_3 \times S^2$  for $u \gg u_{+}$. When $u < u_{+}$, (\ref{xppm}) implies that the brane ends at $u = u_{+}$. To get rid of singularity, changing variables to $u = u_{+} + \rho^2$ with $\rho \ll 1$ gives,
\be
ds^2 = u_{+}^2 (dx \cdot dx)_{2} + \frac{u_{+}^2 (1+q^2)}
{(u_{+}^3 + u_{-}^3)} d\rho^2 + (d\theta_{1}^2 
+ \sin^2 \theta_{1} d\varphi_1^2),
\ee
In fact, one can make another change of variables to the
standard radial variable which corresponds to energy 
scale~\cite{Maldacena:1997re}:
\be \label{u3}
u^3 = U^{-3} - cq + \frac{1}{4}U^3 c^2 (1 + q^2).
\ee
In these new coordinates, the metric takes the form:
\be
ds^2 = (1+ q^2) \frac{dU^2}{U^2} + 
\left(U^{-3} - cq + \frac{1}{4}U^3 c^2 (1 + q^2)\right)^{\frac{2}{3}} (dx \cdot dx)_{3} 
+ (d\theta_{4}^2 + \sin^2 \theta_{4} d\theta_{5}^2). \label{dwm}
\ee
From the roots of $U$,
\be \label{U+}
U_{+} = \left ( \frac{2}{\left | c \right | \sqrt{1+q^2}} \right 
)^{\frac{1}{3}} \le U < \infty.
\ee
Using $U_{+}$ in eqn. (\ref{u3}), one arrives at the original
roots of eqn. (\ref{u3r}).
we notice that the radial coordinate has a range, indicating a
mass gap in the dual theory. We comment more on this in section-IV.\\

\begin{flushleft}
{Branes wrapping CP: {\em Case B}}
\end{flushleft}

\noindent
Let us now discuss embeddings in which the brane wrappings are minimal i.e $\xi= \frac{\pi}{2}$.
 In this case (\ref{xpm}) reduces to
\be
x' = \frac{(qu^3-c)}{u^2 (2cq u^3 - c^2)^{\frac{1}{2}}}.
\ee
It is useful to rescale the parameter $c$ such that $c = C q$;
this removes all $q$ dependence in $x'$:
\be
x' = \frac{(u^3-C)}{u^2 (2C u^3 - C^2)^{\frac{1}{2}}}. \label{dif}
\ee
and the induced metric on the brane is then
\be
ds^2 = \tilde{R}^2 \Big[u^2 ( dx \cdot dx)_2+ \frac{u^4 du^2}{2 C (u^3 - \frac{C}{2})} \Big].
\ee
Note that $x'$ becomes imaginary for  $u^3 < \frac{1}{2} C = u_c^3$, which implies that the brane ends
at $u_{c}$. The induced geometry is non-singular at $u_{c}$
and the embedded hyper surface is  incomplete. \\

\noindent
The induced metric on the $\C\cP^1$ is degenerate. In order to interpret such embeddings in which the $S^2$ is minimal, let us look for D2-brane embedded in the $AdS_4$ and lying at a point in the $\C\cP^3$. In fact the D4-brane  has effectively collapsed to a D2-brane embedded in $AdS_4$, which can be verified deriving explicitly the solution of the D2-brane equations of motion.
The D2-brane field equations in the $AdS_4 \times \C\cP^3 $ target
space are
\be
\frac{1}{3!} \ep^{i_1i_2i_3} f_{i_1 i_2 i_3  m} =e^{-\Phi}\Big[
- \del_{i} (\sqrt{-M} G^{ij} \del_{j} X^{n} g_{mn})
+ \half \sqrt{-M} (G^{ij} \del_{i} X^{n} \del_{j} 
X^{p} g_{np,m})\Big].
\ee
Our ansatz for the worldvolume coordinates is
\be
\eta^{i} = \{ x^0, x^1,  u  \} 
\ee
while the transverse scalars are 
\be
X^{\lambda} = \{ x^2(u) \equiv x(u), \xi, \psi ,\varphi_1,\varphi_2,\theta_1, \theta_2  \}, 
\ee
Then the only equation of motion which is not trivially satisfied by the anatz is
\be
\del_{u} \left ( \frac{u^5 x'}{\sqrt{1+ u^4 (x')^2}} - u^3
\right ) = 0. \label{d3fe}
\ee
Note that this is precisely the $x$ field equation whereas the $u$ field equation follows from it as in the previous case.
Since the general solution of (\ref{d3fe}) is (\ref{dif}), 
this implies that the  D4-brane wrapping
a minimal sphere can be interpreted as a D2-brane. When the flux on the D4-brane is positive we get a D2-brane, whereas negative flux corresponds to anti-D2 brane. These embeddings will be shown to break all
supersymmetry. A new ansatz describing the $D4$-branes wrapping the $\C\cP^1$ and whose worldvolume lie along $x$ is helpful to preserve supersymmetry. Let us consider,
\bea \label{new}
&&\eta^{i} = \{ x^0, x^1, x , \theta_1, \varphi_1 \}; \\ \nonumber
&&X^{\lambda} = \{u, \xi, \psi , \theta_2, \varphi_2 \}; \\ \nonumber
&&F_{\theta_1 \varphi_1} = q \sin \theta_1.
\eea
where all transverse scalars are constant.
Now we have :
\be
\sqrt{-M} =\tilde{R}^3 u^3 L_{(\theta_1 \xi)}
\ee
\be
L_{(\theta_1 \xi)} =\Big(\tilde{R}^4 \cos^4\xi( \sin^2\xi \cos^2\theta_1 + \sin^2\theta_1 ) + q^2 \sin^2\theta_1  \Big)^{\frac{1}{2}}.
\ee
The only field equations which are not already satisfied by the ansatz are 
\bea
u&:& \hspace{5mm} 
u^3 (L_{(\theta_1 \xi)} - q \sin \theta_1) = 0;\\
\xi &:& \hspace{5mm}  
L^{-1}_{(\theta_1 \xi)} u^3 \Big[ \cos^3\xi \sin\xi ( \cos^2\xi \cos^2\theta_1 - 2 \sin^2 \xi\cos^2\theta_1  - 2 \sin^2 \theta_1) \Big] = 0 ;\nonumber
\eea
And the gauge field equation,
\bea
L^{-3}_{(\theta_1 \xi)} u^3 \cos^4\xi \sin^2\xi \cos \theta_1 = 0 \, .
\eea
The  equation deriving from $\theta_1 $ follows from the gauge field equation.
As before the gauge field equation and the the $\xi$ equation  can be solved  either when $(i)$ $\xi= 0 $  or when $(ii)$ $\xi= \frac{\pi}{2} $. In  case $(i)$, 
\be
L_{(\theta_1 \xi)} =(\tilde{R}^4 + q^2)^{\frac{1}{2}} \sin \theta_1 \, ,
\ee
which simplifies the $u$-equation as follows,
\be
u^3 ( (\tilde{R}^4 + q^2)^{\frac{1}{2}} - q ) \sin \theta_1 = 0 \, .
\ee
It has  a solution only for $u=0$ or $ \theta_1 = 0$, which signifies
that the only non-generate solution is  for minimal case. In case $(ii)$,
\be
L_{(\theta_1 \xi)} = q \sin \theta_1
\ee
So the $u$- equation is automatically satisfied. So there exists a solution  with non-zero flux $q$ for any $u_0$.
We use kappa symmetry projections to check the supersymmetry preserved by these embeddings in the next section. It
should be mentioned that kappa symmetry gauge fixing of the $D4$ brane embeddings considered in this paper from the world volume point of view is a subject which requires further study. For instance, for the case of fundamental strings and D2 branes wrapping 
$AdS_2 \times S^1$ and at the minkowski boundary of $AdS_4$ interesting gauges were discussed 
in~\cite{Grassi:2009yj}.

\section{Supersymmetry of Embeddings}\label{Ap:Killing}
\noindent

In this section, we check the supersymmetry preserved by our brane solutions corresponding to cases A and B, found in the last section. Most of the aspects
of killing spinors are presented in the
appendix B and an analysis related to supersymmetric
and non-supersymmetric defects is given below. 

The supersymmetry of the $AdS_3 \times 
CP^1$ brane embeddings discussed in the previous section is checked as follows. The explicit form of the kappa symmetry projection, $\Gamma \ep = \ep$ (see appendix B) is:
\be \label{kappacon}
\ep = -\frac{1}{u^3 (1+q^2)} \g^{01}  
\left ( (qu^3-c) \g^2 + \sqrt{(u^6 + 2 c q u^3 - c^2)} \g^3 \right ) 
\left ( -\g_{58\natural} + q \right)\ep \, , 
\ee
where $\ep$ is the Killing spinor corresponding to the unbroken supersymmetry. 
Preservation of supersymmetry requires that this condition must
be satisfied for some subset of the background Killing spinors
at all points on the brane worldvolume. In particular, it
must hold at all values of $x^p = (x^0,x^1,x^2)$. Below, we explicitly calculate the supersymmetry preserved
by the $c=0$ embeddings and argue that $c \neq 0$ ones break all supersymmetry.
Before proceeding, it is necessary to project the solutions for the killing spinors on to the
brane world volume. This is done as follows. 
Using the change of coordinates $u = e^r$ in eqn.(\ref{ads4}), the killing spinor equation given in
eqn. (\ref{kill2}) takes the form:
\be
\partial_{r}\epsilon -
\frac{1}{2} \hat \g \, \gamma_{r}\epsilon=0\, , \quad 
\partial_{p}\epsilon +
\frac{1}{2} \gamma_p(\hat \g + \gamma_{r})\epsilon=0\, ,
\ee
where $p$ runs over $0,1,2$. The solution of these equations is:
\be
\epsilon=e^{\frac{1}{2} r  \hat \g \gamma_r} \Big( 1+\frac{1}{2}\, \hat \g x^p \gamma_p
(1- \hat \g \gamma_r)\Big)\epsilon_0\ ,
\ee
where $\epsilon_0$ is an arbitrary constant spinor. 
\noindent

Using the solution in eqn. (\ref{xq0}) and after a few manipulations, it is convenient to write the killing spinors as:
\be \label{kill}
\epsilon= \Big( e^{\frac{1}{2} r  \hat \g \gamma_r} + \frac{1}{2}e^{\frac{1}{2} r  \hat \g \gamma_r} \, 
(\g_r + \hat \g) (x^l \gamma_l + \tilde x \gamma_2)
- \frac{q}{2} e^{-r}(\g_r + \hat \g) \g_2 \Big)\,  h(\C\cP^3)\, \epsilon_0\ ,
\ee
where $l=0,1$, and $h(\C\cP^3)$ is given explicitly in
eqn.(\ref{killcp}). From (\ref{kappacon}) and (\ref{kill}), noting the terms in
the Killing spinors which are linear in $x^l$, we find
the following condition:
\bea \label{k1}
&&e^{\frac{1}{2} r  \hat \g \gamma_r}(\g_r + \hat \g)\,\g_l\, h(\C\cP^3)\, \epsilon_0 \nonumber \\
&&=  - \frac{1}{ (1+q^2)} \g_{01}  \,\Big( e^{-\frac{1}{2} r  \hat \g \gamma_r} q( \g_3 - \g_{258\natural}) +
e^{\frac{1}{2} r  \hat \g \gamma_r} (-\g_{358\natural} + q^2 \g_2)
\Big)\, (\g_r + \hat \g)\,\g_l\, h(\C\cP^3)\, \epsilon_0 \, .
\eea
To proceed further, we start by noticing that terms in the function $h(\C\cP^3)$ in (\ref{killcp}), which do not contain gamma 
matrices cancel generally. For the rest of the terms in  $h(\C\cP^3)$, one checks that $\Gamma$ commutes
with $\hat\g \g_5$ and $\g_{8\natural}$. Further, one can multiply by the inverse of $h(\C\cP^3)$ as,
\be
{h(\C\cP^3)}^{-1} (1- \Gamma) h(\C\cP^3) \epsilon_0 = 0 \, ,
\ee
to eliminate $h(\C\cP^3)$ completely from the equations. The eqn. (\ref{k1}) gives two sets of conditions:
\be \label{k4a}
 \frac{1}{ (1+q^2)} \g_{01} (\g_3 + \g_{258\natural}) (\g_r + \hat \g) \epsilon_0= 0 \, , 
\ee
and 
\be \label{k4b}
 \frac{1}{ (1+q^2)} \g_{01}  ( \g_{358\natural} + q^2 \g_2)(\g_r + \hat \g)\epsilon_0  = (\g_r + \hat \g)\epsilon_0 \, .
\ee
To solve eqns. (\ref{k4a}) and  (\ref{k4b}), let us introduce the projections,
\be \label{proj}
 \g_{012} \epsilon_0 = \pm \epsilon_0^{\pm} \, .
\ee
Using eqn. (\ref{proj}), both the conditions (\ref{k4a}) and  (\ref{k4b}) are solved by introducing the
projection,
\be \label{r1}
(1 + \g_{3258\natural})\, \epsilon_0^{+} = 0 \,
\ee
where, $\epsilon_0^{-}$ disappears from the equations. Thus, if one introduces the projections
$\g_{3258\natural} \epsilon_0^{+} = \pm \eta_{\pm}$, then eqn. (\ref{r1}) means that $\eta_+$ 
is eliminated and $\eta_-$ remains undetermined at this level.  Eqn. (\ref{r1}) will be analyzed further below. \\

\noindent
The remaining terms from (\ref{kill}) give the following kappa symmetry projection:
\bea \label{k2}
&&\left[e^{\frac{1}{2} r  \hat \g \gamma_r}( 1 + \frac{1}{2}\tilde x (\g_r + \hat \g) \g_2 ) 
- e^{-\frac{1}{2} r  \hat \g \gamma_r}\frac{q}{2}(\g_r + \hat \g) \g_2\right]\, h(\C\cP^3)\, \epsilon_0 \\ \nonumber 
&&= - \frac{1}{ (1+q^2)} \g_{01} \left[ \{e^{-\frac{1}{2} r  \hat \g \gamma_r} q( \g_3 - \g_{258\natural})
+ e^{\frac{1}{2} r  \hat \g \gamma_r} (-\g_{358\natural} + q^2 \g_2)\} ( 1 + \frac{1}{2}\tilde x (\g_r + \hat \g) \g_2 ) \right. \\ \nonumber
&& \left. - \{ e^{\frac{1}{2} r  \hat \g \gamma_r} q( \g_3 - \g_{258\natural}) + e^{-\frac{1}{2} r  \hat \g \gamma_r} (-\g_{358\natural} + q^2 \g_2)\} \frac{q}{2}(\g_r + \hat \g) \g_2\right]\, h(\C\cP^3)\, \epsilon_0 \, .
\eea
The kappa-symmetry projection condition (\ref{kappacon}) has to be satisfied at all points on the
brane world-volume. Commuting  $h(\C\cP^3)$ as before, thus eqn. (\ref{k2}) gives rise to two independent conditions,
\bea \label{k3a}
&&\epsilon_0^{-} + ( 1 + \tilde x \g_{32} )\epsilon_0^{+}  \\ \nonumber
&&= - \frac{1}{ (1+q^2)} \g_{01} \left((-\g_{358\natural} + q^2 \g_2) (\epsilon_0^{-} + ( 1 + \tilde x \g_{32} )\epsilon_0^{+}) - q^2  (\g_3 - \g_{258\natural}) \g_{32} \epsilon_0^{+}
\right)
\eea
and
\be \label{k3b}
q \g_{32} \epsilon_0^{+} = \frac{1}{ (1+q^2)} \g_{01} \left( q  (\g_3 - \g_{258\natural})(\epsilon_0^{-} + ( 1 + \tilde x \g_{32} )\epsilon_0^{+}) - q (-\g_{358\natural} + q^2 \g_2) \g_{32} \epsilon_0^{+}
\right)
\ee
If one introduces the projections,
\be \label{r2}
 \g_{3258\natural}\, \epsilon_0^{-} = \pm \lambda_{\pm} \, ,
\ee
then both the eqns. (\ref{k3a}) and (\ref{k3b}) can be solved by using the relation,
\be
2 \lambda_{-} = - \tilde x\, \g_{32} \eta_- \, .
\ee
$\lambda_+$ is undetermined as well.  Now, let us analyze the conditions (\ref{r1})
and (\ref{r2}) further. If  $s_1 = s_2$, then
\be \label{s}
\g_{58\natural} \epsilon_0 = \hat \g \epsilon_0 \, ,
\ee
turning the projection condition (\ref{r1}) in to,
\be \label{r3}
\g_{01} \epsilon_0 =   \epsilon_0 \, .
\ee
On the other hand, if $s_1 = - s_2$, the condition is,
\be \label{r4}
\g_{01} \epsilon_0 =  - \epsilon_0 \, .
\ee
Notice that $s_1 = s_2$ is satisfied by eight supersymmetries and they satisfy the projection in eqn. (\ref{r3}).
When $s_1 = - s_2$, there are sixteen supersymmetries and they satisfy the projection conditions (\ref{r4}).
Thus, one sees that twelve of the supersymmetries are 
preserved and half of the target space supersymmetry is broken by these embeddings. Thus, the projections do not depend on the flux, but, on the value of $\tilde x$. Thus,
each extra defect breaks the supersymmetry further. \\

\noindent
For $c \neq 0$, one has different type of conditions. For instance, one ends up with conditions of the form:
\be
 -\frac{1}{u^3 (1+q^2)} \g^{01}  
\left ( - (qu^3-c) \g^{258\natural} + q \sqrt{(u^6 + 2 c q u^3 - c^2)} \g^3 \right )h(\C\cP^3) \, \epsilon_0 = 0
\ee
From inspection, one sees that these conditions cannot hold at all points on the brane world volume. Thus,
the $c\neq 0$ embeddings corresponding to asymptotically $AdS_3 \times S^2$ branes, break all supersymmetry. \\
\noindent
For the embeddings corresponding to {\em case B}, the kappa-symmetry projection leads to:
\be
\epsilon = - \g_{01}\Big( \frac{\sqrt{2Cu^3 - C^2}}{u^3} \g_3 + (u^3-C)\g_2  \Big) \epsilon
\ee
Using the form of the killing spinors in eqn. (\ref{kill}), we end up with the following conditions:
\be
\frac{\sqrt{2Cu^3 -C^2}}{u^3} \g_{013} \, h(\C\cP^3) \, \epsilon_0 = 0\, , \qquad
(u^3 - C)\g_{012} h(\C\cP^3) \, \epsilon_0 = 0
\ee
Clearly, these conditions cannot be satisfied for any non zero $C$ and  $\epsilon_0$. Thus, there are no
non zero solutions to the kappa symmetry projections on the world volume and hence these embeddings break
all supersymmetry as well, as in the cases considered in~\cite{Skenderis:2002vf}. One attributes the breaking of all supersymmetries to the misaligned branes in the $AdS_4\times \C\cP^3$ background. For instance, if one had considered an embedding of the form given in eqn. (\ref{new}),
the induced metric in this case is very simple:
\be
ds^2 = u^2 (dx.dx) \, .
\ee
In this case, the kappa symmetry projection is straightforward, giving:
\be \label{sgn}
\ep = - sgn(q) \, \g_{012} \ep \, .
\ee
Note that for the sign of $\Gamma$ we have chosen, only embeddings which satisfy $(1-\Gamma)\epsilon = 0$ respect supersymmetry and those with $(1+\Gamma)\epsilon = 0$ break all supersymmetries. From 
eqn. (\ref{sgn}), one
can check that, for the embeddings with $q = +1$, the condition in eqn. (\ref{r1}) is enough and hence, half of the supersymmetry is preserved. The ones with $q = -1$ break
all supersymmetry, as there is an over all negative sign, incompatible with our choice of $\Gamma$.
Although, one expects that in the plane wave limit, both the signs of charges are compatible with
the kappa symmetry projection. It would be interesting to further the analysis of this work to 
the plane wave limit and study the possible brane embeddings in that context.
\\
\section{Field Theory Analysis} \label{fieldtheory}
In this section, we discuss the field theories dual to the brane
embeddings constructed in section-II. For a full review of the Lagrangian and symmetries of the ABJM theories, we 
refer to~\cite{Aharony:2008ug} and collect a few facts below, needed later on.\\

The ABJM theory is a $2+1$ dimensional, $U(N_c)_{k} \times U(N_c)_{-k}$ gauge theory with a Chern-Simons term for each gauge group factor. The two Chern-Simons terms have equal but opposite levels, $k$ and $-k$. The field content consists of ${\mathcal N} =2$ vector superfields $V_i$ and adjoint chiral superfields $\Phi_i$, where $i=1,2$ labels the each $U(N_c)$ factor. There are also four ${\mathcal N} =2$ chiral superfields, $A_1$, $A_2$, $B_1$ and $B_2$, where $A_1$ and $A_2$ are in the bifundamental $(N_c,\overline{N_c})$ representation and the $B_1$ and $B_2$ are in the anti-bifundamental $(\overline{N_c},N_c)$ representation. In 
particular, we note that the chiral superfields $\phi_i$ act
as constraints and removing them by their equations of motion leads
to the ABJM superpotential:
\be
\label{abjmsupo2}
W_{\text{ABJM}} = \frac{2 \pi}{k} \varepsilon^{ab} \, \varepsilon^{\dot{a} \dot{b}} \, \Tr \left( A_a B_{\dot{a}} A_b B_{\dot{b}} \right),
\ee
which exhibits an  $SU(2)_A \times SU(2)_B$ acting separately on $A_a$ and $B_{\dot{a}}$ respectively. As discussed in~\cite{Aharony:2008ug}, the R-symmetry of the theory, $SO(3)_R \equiv SU(2)_R$, does not commute with these, the full R-symmetry is $SU(4)_R \equiv SO(6)_R$, and the theory is in fact ${\mathcal N} =6$ supersymmetric. There is also a $U(1)_b$ ``baryon number'' symmetry under which $A_i \rightarrow e^{i \alpha} A_i$ and $B_i \rightarrow e^{-i \alpha} B_i$, which will be useful in analysis of brane embeddings from M-theory point of view.\\

Brane embeddings in section-\ref{dbranembd}, have been useful for a general analysis of adding flavor to ABJM theories~\cite{Ammon:2009wc}, where both the
type IIB and M-theory point of view were presented.  Below, we rely on the analysis of~\cite{Ammon:2009wc}, to provide a discussion of defect conformal field theories for the present case.
Although, our brane set ups appear different from the point of view of IIA superstrings, they also satisfy the same consistency conditions of an $M5$-brane embedding in M-theory~\cite{Ammon:2009wc} and hence are $SU(4)$ equivalent. We comment on this below and also on the construction of operators in the dCFT's dual to $AdS_4 \times \C\cP^3$, in comparison to the case of $AdS_5 \times S^5$. For instance, in the type IIB case, one
can start with a $D3-D5$ branes and take the near horizon limit on the $D3$-branes to obtain an embedding of a probe $D5$-brane in the
$AdS_5 \times S^5$ geometry. It is possible in this way to analyze the meson spectrum when the branes are separated and/or are top of
each other~\cite{Myers:2006qr}. Further, one can also get insights in to the couplings of various fields localized on the
intersection of original D3-D5 branes~\cite{ooguri1}. In the case of type IIA on $AdS_4 \times \C\cP^3$, the corresponding way to analyze, say, the D2-D4 defect CFT's, is again to start from a
D3-D3 color and flavor D-brane system in type IIB and trace the embeddings through a series
of transformations to come back to type IIA. For the case of interest corresponding to the embeddings in
section-II, it is useful
to start with the following type IIB brane set up~\cite{Ammon:2009wc}: 
\begin{center}
        \begin{tabular}{|c|cccccccccc|}\hline
                &0&1&2&3&4&5&6&7&8&9\\ \hline
NS5&$\bullet$&$\bullet$&$\bullet$&$\bullet$&$\bullet$&$\bullet$&--&--&--&--\\
$(1,k)$5&$\bullet$&$\bullet$&$\bullet$&$[3,7]_\theta$&$[4,8]_\theta$&$[5,9]_\theta$&--&--&--&--\\
$N_c$ D3&$\bullet$&$\bullet$&$\bullet$&--&--&--&$\bullet$&--&--&--\\\hline
        \end{tabular}
\end{center}
obtained after a series of transformations from the set up in~\cite{Aharony:2008ug}. The direction $x^6$ is a circle. 
In the original configuration of~\cite{Aharony:2008ug},
there were two stacks of NS5- and NS5$^{\prime}$-branes separated in the $x^6$ direction. The $N_c$ D3-branes actually break on the NS5-branes and also the $k$ D5-branes were coincident to the the NS5$^{\prime}$-brane	 in $x^6$. The
$(1,k)5$-brane comes about by binding the $k$ D5-branes to the NS5$^{\prime}$-brane tilted at an angle $\theta$  in the $(37)$, $(48)$ and $(59)$ plane denoted respectively as
$[3,7]_\theta$, $[4,8]_\theta$ and $[5,9]_\theta$. The $D4$-brane embeddings can now be understood from above construction as follows. One adds two stacks of D3-branes, each with $N_f$ coincident D3-branes, on opposite sides of $x^6$ direction. T-dualizing along $x^6$, 
the D3-branes become D4-branes and the gauge group is now
$U(N_f) \times U(N_f)$ (due to a $Z_2$ valued wilson line, see~\cite{Hikida:2009tp}). On the field theory side, the
additional $N_f$ branes correspond to co-dimension one non-chiral flavor,
i.e., $(1+1)$-dimensional defect in the ambient $(2+1)$-dimensional ABJM theory (see~\cite{Ammon:2009wc}). Recall that if the D3/Dp intersection has 4 Neumann-Dirichlet (ND) directions then the corresponding flavor fields (from 3-p and p-3 strings) will produce non-chiral flavor, simply because the fields are arranged in hypermultiplets, whereas with 8 ND directions we can obtain chiral flavor, as occurs for the 8 ND D3/D7 intersection~\cite{Harvey:2007ab,Buchbinder:2007ar,Harvey:2008zz}. The flavor fields will have ${\mathcal N}=(4,4)$ supersymmetry broken to ${\mathcal N}=(3,3)$ supersymmetry when the Chern-Simons level $k\geq2$. \\ 

\noindent

Before proceeding, let us note that brane constructions involving
D2-D4 system and D2-D8 systems were proposed in~\cite{Fujita:2009kw}, to give a holographic description of 
Fractional Quantum Hall Effect. For our alignment of
branes, both supersymmetric and
non-supersymmetric  
solutions were presented in eqn. (\ref{xpm}) at different values of $\xi =0$ in eqn. (\ref{xppm}) and
$\xi = \pi/2$ in (\ref{dif}).  These solutions were in fact
used in section-IV to explicitly check the kappa symmetry conditions. The  $\xi =0$ solution corresponding to a $D4$-branes wrapping $\C\cP^1$ in
$\theta_1,\phi_1$ directions appeared in~\cite{Fujita:2009kw}
in their construction of edge states. We also note that the brane set up in~\cite{Fujita:2009kw} corresponds to matter fields coupled to single gauge group and were parity breaking, leading to domain wall type configurations, reminiscent of fractional $M2$-branes~\cite{Aharony:2008gk}.
In the present case, as discussed above, strings from color to flavor D3-branes introduce flavor in both gauge groups. \\

The existence of these solutions, in particular, of the form 
presented in eqn.(\ref{xq0}) is in agreement with the AdS/CFT
result regarding the dimension of scalar fields in $AdS_4$, i.e.,
the boundary value of the scalar field may be identified with
the source of the gauge theory-operator $\mathcal O$, and 
$\left< \mathcal O \right>$ is the vev (vacuum expectation value) of $\mathcal O$~\cite{Maldacena:1997re}. Now, the active scalar in our embeddings behaves at large $u$ as:
\be \label{active}
u x \approx u x_0 + \frac{c}{4u^3} \, ,
\ee
one expects that the scalar is dual to an operator $O_{x}$ of conformal dimension $3$ in the defect theory. Let us discuss some of
the difficulties involved with the identification of the dual operator in the present
case. As already mentioned, our D4-brane at $\xi =0$ preserves an $SU(2)_V \times SU(2)_H \times U(1)$ symmetry of
the original $\C\cP^3$ geometry.  The
$SU(2)_H$ rotates $(\theta_1, \varphi_1)$ and corresponds to
the symmetry of the $(1+1)$-dimensional hypermultiplet. The
$SU(2)_V$ comes from the symmetries of the transverse directions
and contains scalar fields of ambient vector multiplet, restricted to the defect theory. It is 
nice to compare this with the M-theory analysis of~\cite{Ammon:2009wc}, where the R-symmetry preserved by 
$M5$-brane set up was argued to be $SU(2) \times SU(2) \times U(1) \times U(1)_b$. Our D4-branes naturally satisfy the 
consistency conditions of the $M5$-branes.
This comes about by noting that the $\xi =0$ case of the 
embedding discussed in eqn. (\ref{xppm}), upon uplifting to M-theory corresponds to an $M5$-brane embedding given by the
equations $z_3 =z_4 =0$, in terms of the coordinates given before
eqn. (\ref{S7-form}). Thus, the $SU(4)$ symmetry of $M5$-branes
breaks to $SU(2) \times SU(2) \times U(1)$. The additional $U(1)_b$
is common phase shift of all $z$ coordinates corresponding to a
shift in the M-theory circle. Now, the operator dual to the scalar field $ux$ has to be constructed out of the
defect flavor scalars and scalar components of the superfields $A_a$ and $B_a$, restricted to the defect. Let us denote the defect flavor scalars as $q_i^n$, where $i=1,2$ labels the two gauge groups and $n=1,2$ labels the two complex scalars of an $N=(4,4)$ hypermultiplet. Following the analysis of~\cite{ooguri1,johanna,Ammon:2009wc}, the 
operator is expected to be formed out of two parts, a part coming from the scalars $A_a$ and $B_a$, which have dimension $1/2$ and
another coming from $q_i^n$, which are dimension zero. The
defect CFT
operator is expected to be a space-time scalar and an R-singlet,
and this is in accordance with the fact that the
$\C\cP^1$ part of our embedding in eqn. (\ref{d4ads3s2}) is undeformed. This situation is to be contrasted this with the case of massive flavor studied in~\cite{Jensen:2010vx}, where the
non-trivial 3-cycle $\bR P^3$ that the D6-brane wraps, is 
deformed as a function of the radial coordinates. Since,
the defect scalars $q_i^n$ are inert under all four global groups, from symmetry constraints they form a singlet of 
$SU(2) \times SU(2)$, as $\bar q q$, with all indices contracted.
As in~\cite{ooguri1}, operators with higher dimension can be
formed by constructing $SU(2) \times SU(2)$ singlets of the
the scalars $A_a$ and $B_a$. For the present case, these can be constructed as in~\cite{Ammon:2009wc}, by forming fields 
$C =\left(\begin{array}{c} A_1 \\ B_1^* \end{array} \right), \qquad D = \left(\begin{array}{c} B_2^* \\ A_2 \end{array} \right) \, ,$ which are doublets under $SU(2)_1 \times SU(2)_2$ respectively. The fields $C,D$ carry charges $+1,-1$ under the $U(1)_D$ respectively. 
Also the ABJM baryon number $U(1)_b$, acts as $A_a \ra e^{i\a} A_a$ and $B_a \ra e^{-i\a} B_a$. Using
these properties, one can form a dimension one operator, which takes a general form ${\mathcal O}_H \approx \bar q q \bar X_H X_H$, where $X_H$ is to be determined in terms of $C$ or $D$. 
The defect CFT operator ${\mathcal O}$ of dimension three, is expected to
be a supercharge descendant of ${\mathcal O}_H$. We stress that, at this stage,
this is an assumption, based on symmetry arguments and 
it is necessary to learn about the full coupling of 3d CFT with 2d defect fields to identify it.  The full low energy theory on the dCFT's, which contains some of the possible terms discussed
in~\cite{Ammon:2009wc} is also needed. This discussion is also to be contrasted with the construction of defect operators discussed in the context of D3-D5 theories which are of dimension 
four~\cite{ooguri1,johanna}. In that case, the operator dual to the
scalar field was a certain four supercharge descendant of the
dimension two chiral primary on the defect~\cite{ooguri1,Skenderis:2002vf}. The knowledge of spectrum of open and closed strings modes coming from the Kaluza-Klein reduction of D4-branes on $\C\cP^1$ is useful to
understand chiral primaries of defect CFT. In this context, the fluctuation analysis of $1+1$ defect CFT's originating from branes in M-theory are useful~\cite{Fiol:2010un}. We are currently investigating these questions.\\

\noindent

We note that $x_0$ has the interpretation of a source for an operator ${\mathcal O}_x$, with $c$ as its expectation value.
The operators discussed above correspond to the lowest components of 2d superfield and giving vev to them, breaks supersymmetry, which we explicitly confirmed in section-III.
This provides evidence that a full dictionary could be developed between the defect CFT operators and objects in the
bulk. It might be possible to develop such a dictionary more readily in the plane wave limit, owing to the exact solvability of the string spectrum~\cite{Lee:2002cu}. 
Since, the D4-brane solutions used in this paper do not back react on the bulk,
there is a possibility in which the boundary theory remains
conformal, but the defect theory runs. This can be inferred from the asymptotically $AdS_3 \times \C\cP^1$ embeddings in
to the ambient $AdS_4 \times \C\cP^3$ background given in
eqn. (\ref{dwm}). In fact, this RG flow is driven by the
vev of our active scalar in (\ref{active}). One feature of
this RG flow is the mass gap observed in the radial coordinate
$U$ in eqn. (\ref{U+}). To check this, one needs to study the
embeddings of D4-branes not just in the near horizon limit, but the full ten dimensional background, and the related embeddings
of probe $M5$-branes in the geometry of 11d supergravity backgrounds~\cite{Yamaguchi:2003ay,Lunin:2007ab}. In this case, $x$ is a general function $x(u)$ satisfying certain embedding equations and its profile could show 
interesting features. As 
we know from the near horizon analysis done in this paper, such embeddings are not expected to be supersymmetric unless
$x$ is a constant.

\section{Discussion}\label{conclusion}


In this paper, we discussed various brane embeddings in $AdS_4\times \C\cP^3$ and the supersymmetry preserved 
by them. In general AdS/dCFT dualities, especially in
the case of duality between Type IIB string theory on $AdS_5 \times S^5$ and $N=4$ super Yang-Mils system, there exists a good notion of renormalization group flow on the defect CFT, coming from
the embedded branes.  In the present case, we argued that the defect theory should be associated with the boundary of the $AdS_3$ of the $AdS_3 \times \C\cP^1$ D4-brane. We expect such RG flows to hold for the present case as well, when the defect theory
is conformal.  So far we considered a single $D4$-brane, whose back reaction on the near-horizon geometry
can be neglected in the 't Hooft limit, allowing it to be treated as a probe hosting open
strings. Thus, the deformations in which the boundary theory
remains conformal but the defect theory runs are possible. An example is the
asymptotically $AdS_3 \times \C\cP^1$ embeddings discussed in
section-II.  It would be interesting to consider the
back reaction of the $D4$-brane in the target space geometry. In this case, one also has to find fully localized brane solutions in $AdS \times CP$ backgrounds. A fully localized solution is also important to understand the back reaction effects of the embeddings considered.  This would also give insight in to the
the dynamics of boundary versus bulk modes in the CFT coming from gravity fluctuations.
Further, intersecting brane solutions in $AdS_4\times \C\cP^3$ backgrounds could be constructed to get 
insights in to the spectrum of defect conformal field 
theory~\cite{Fujita:2009kw}. It is an interesting exercise to construct
the defect CFT operators and understand the correlation functions along the lines of~\cite{ooguri1,Sethi:1997zz}-\cite{Hori:2000ic}.\\

\noindent

Although we considered specific embeddings in the 
$AdS_4\times \C\cP^3$ background, it is possible to consider and classify other embeddings. In this context, it would be 
useful to explore the possible branes in the Penrose limits
of $AdS_4\times \C\cP^3$~\cite{Nishioka:2008gz}. We expect that the structure of branes in the Penrose limit is quite
rich with branes longitudinal or transverse to light-cone directions and also instantonic branes which can come from euclidean branes in AdS. The classification of embeddings can be done following the analysis in~\cite{Skenderis:2002vf},
using the number of intersection directions of Neumann and Dirichlet directions.
Supersymmetry preserved by the brane configurations depends on the coordinate splitting. When they are
sitting at arbitrary positions of the transverse space they preserve one quarter of
supersymmetry, but they preserve twelve supercharges when located at the origin of the transverse
space. \\

In Section-II, we noticed that some brane configurations
were stable mainly because the mass of the tachyonic mode was below the Breitenlohner-Freedman bound in AdS, but this need not be so. It is also interesting to consider other kind of solutions corresponding to certain applications to condensed matter systems, where 
the addition of charge density and/or magnetic field is
important~\cite{Hartnoll:2007ai}. Then, phase transitions shown by many lower dimensional systems can be holographically captured,
by studying $Dp$-branes in the $ AdS_4\times \C\cP^3$
geometry, both at zero temperature and finite temperature.
For instance, one can turn on density corresponding to the flavor charge coupled to a magnetic field, as $F = A'_t(r)dt\wedge dr+B dx^1\wedge dx^2$.
At zero temperature, novel holographic Berezinskii-Kosterlitz-Thouless (BKT) transitions have been identified in~\cite{Jensen:2010ga,Jensen:2010vx} corresponding to configurations where the Breitenlohner-Freedman bound~\cite{Breitenlohner:1982jf}
is violated in a controlled setting. A construction of phase diagram~\cite{Evans:2010hi} of embeddings of $Dp$-branes constructed in this work, in the near horizon, as well as in the decompactified limit in M-theory, should be an interesting exercise.

\noindent
\acknowledgments
We thank Alok Kumar for useful discussions and the organizers of String Theory and Fundamental Physics, Kanha, where some part of this work was done. B.C. thanks IFT, Sao Paulo for support, IOP, Bhubaneswar, HRI, Allahabad and MPI, Halle, for hospitality, during the course of this work. B.C. is supported by the CNPQ grant 150417/2009-8. B.C. would also like to thank D. Sorokin for useful discussions. B. P. would like to thank the Theory Division of CERN physics department for their hospitalities during the course of the work. The work of B. P. is supported by the European Commission under the ERC Advanced Grant $226371$. 

\medskip

\appendix

\section{D-brane Field equations}
In this appendix, we summarize the D-brane field equations derived in all generality and for all Dp-branes in~\cite{Skenderis:2002vf}. 
The worldvolume action for a single $Dp$-brane is given by
\bea
I_{p} &=& I_{DBI} +  I_{WZ}  \label{dbiact} \\
I_{DBI} &=&
=- T_{p} \int_{M} d^{p+1}\eta e^{-\Phi} 
\sqrt{-\det \left (g_{ij} + {\cal{F}}_{ij} \right )}, \nonumber \qquad
I_{WZ} = T_{p} \int_{M} e^{\cal{F}} \wedge C, \nonumber
\eea
with $T_p$ the Dp-brane tension. 
Here $\eta^i$ are the coordinates of the $(p+1)$-dimensional
worldvolume with metric $g_{ij} = g_{mn} \del_{i} X^{m} \del_{j} X^{n}$, following from
the space-time string frame metric $g_{mn}$. Worldvolume field strength ${\cal{F}} = F - B$ is the gauge invariant two-form with 
$B_{ij}=\del_i X^m \del_j X^n B_{mn}$ the pullback of the target 
space NS-NS 2-form.
Finally, we summarize the D-brane field equations
\bea
&& \del_{i} (e^{-\Phi} \sqrt{-M} \theta^{ii_1}) =  
 \ep^{i_1..i_{p+1}} \sum_{n \ge 0} \frac{1}{n! (2!)^n (q-1)!} 
({\cal F})^{n}_{i_2...i_{2n+1}} 
\bar{F}_{i_{2n+2}... i_{p+1}}. \label{finfe} \\
&&\hspace{-10mm}\sum_{n \ge 0} \frac{1}{n! (2!)^n q!} 
\ep^{i_1..i_{p+1}}
({\cal F})^{n}_{i_1..i_{2n}} \bar{F}_{m i_{2n+1}... i_{p+1}}=
e^{-\Phi} \left( 
\sqrt{-M} \left ( G^{ij} \del_{i} X^{p} \del_{j} X^{n} g_{mn} 
\Phi_{,p} - \Phi_{,m} \right ) - {\cal K}_m \right). \nonumber 
\eea
where 
\be
{\cal K}_m=-\del_{i} (\sqrt{-M} G^{ij}) \del_{j} X^{n} g_{mn} 
-\sqrt{-M} M^{ij} \left ( (\del_{i} \del_{j} X^{n}) g_{mn} 
   + \tilde{\Gamma}_{mnp} \del_{i} X^{n} \del_{j} X^{p} \right ) 
\ee
\be
\bar{F}_{m_1..m_{q+1}} = f_{m_1..m_{q+1}} - \frac{(q+1)!}{3! (q-2)!}
H_{[m_1..m_3} C_{m_4..m_{q+1}]}.
\label{rrfe}
\ee
where the following notations are introduced,
\be
M_{ij} = (\del_{i} X^{m} \del_{j} X^{n} g_{mn} - 
\del_{i} X^{m} \del_{j} X^{n} B_{mn} + F_{ij}).
\ee
and the inverse as $M^{ij}$ such that $M^{ij} M_{jk} = \delta^{i}_{\hspace{2mm} k}$.
Moreover, $G^{ij} \equiv M^{(ij)}$ 
and $\theta^{ij} \equiv M^{[ij]}$.
And we have $\tilde{\Gamma} = \G - \half H$, with  $\G_{mnp}$ is the Levi-Civita connection of the target space
metric and $H_{mnp}$ is the field strength of the
NS-NS two form.\\
For the special case, when $F_{ij} = B_{mn} = \Phi = 0$, one has,
\be
J^{m}=- \sqrt{-g} g^{ij} {\cal K}^{m}_{ij}
\ee
where 
\be \label{secondf}
{\cal K}^{m}_{ij}=\gamma^{k}_{ij} \del_{k} X^{m}
- (\del_{i} \del_{j} X^{m}) - \Gamma^{m}_{np} \del_{i}X^{n} \del_{j} X^{p}
\ee
is the second fundamental form ($\gamma^k_{ij}$ is the Levi-Civita 
connection of the induced worldvolume metric).
If in addition $J_{m} = 0$, the field equation becomes
\be
g^{ij} {\cal K}^{m}_{ij} = 0, \label{kcon}
\ee
that is, the trace of the second fundamental form of the embedding
must be zero.


\section{Killing Spinors of $\C\cP^3$ in the basis of section-\ref{Ap:Killing} }


\noindent

In this appendix, we provide the necessary details to analyze the BPS configurations of D-branes that are wrapped on compact
portions of our background, and are point like in the AdS. In order for the D-brane to be
supersymmetric, we only need to check that the kappa-symmetry conditions \cite{Cederwall:1996pv,Aganagic:1996pe,Cederwall:1996ri,
Bergshoeff:1996tu, Aganagic:1996nn, Bergshoeff:1997kr} 
\be
\Gamma \ep = \ep. \label{projc}
\ee
is satisfied, where $\ep$ is the Killing spinor corresponding to the unbroken supersymmetry. As in~\cite{Skenderis:2002vf} the projector $\Gamma$ involving the flux on the worldvolume of the brane is considered for convenience.
The projection matrix is
given by
\be
d^{p+1} \eta \Gamma = - e^{-\Phi} {\cal{L}}_{DBI}^{-1} 
e^{\cal F} \wedge X |_{vol}, \label{kap1} 
\ee
with $X = \bigoplus_{n} \G_{(2n+1)} \G_{11}^{n+1} \id $ and ${\cal{L}}_{DBI}^{-1}$  is to be 
evaluated on the background. Also, $\G_{(n)} = \frac{1}{n!} d\eta^{i_{n}} \wedge ... \wedge d\eta^{i_1} 
\G_{i_1...i_n} $ where $\G_{i_1...i_n}$ is the pullback for the target space
gamma matrices $\G_{i_1...i_n} =\del_{i_1} X^{m_1} ... \del_{i_n} X^{m_n} \G_{m_1...m_n}$.
$\G$ has the special property that it squares to one and 
is traceless. It follows that one can use $\G$ to project out 
half of the worldvolume fermions, thus preserving supersymmetry.
In type IIA case, the two 16-component Majorana-Weyl spinorial coordinates of opposite
chirality form a 32-component Majorana spinor $\psi^{\alpha}$, $\alpha = 1, .., 32$. 
In the Weyl basis, the corresponding gamma matrices are given by real block-off-diagonal matrices with a diagonal $\G_{11}$-matrix.
To find the relevant Killing spinor equation for this background we look at 
the supersymmetry transformation of the gravitino
\be
\delta\Psi_\mu=D_\mu\epsilon-\frac{1}{288}\Big(
\Gamma_\mu^{\,\n\l\r\s}
-8\delta_\mu^\nu\Gamma^{\lambda\r\sigma}\Big)F_{\n\l\r\s}\epsilon\,,\qquad
D_\mu\epsilon=\partial_\mu\epsilon+\frac{1}{4}\omega_\mu^{ab}\gamma_{ab}\epsilon\,.
\ee
The 4-form corresponding to the $AdS_4 \times S^7$ solution is
$F_{\nu\lambda\rho\sigma} = 6\, \varepsilon_{\nu\lambda\rho\sigma}$, 
where the epsilon symbol is the volume form on $AdS_4$ (so the indices take 
the values $0,1,2,3$). Plugging this into the variation
above one finds the Killing spinor equation
\be \label{kill2}
D_\mu \epsilon = \frac{1}{2} \hat \gamma \gamma_\mu \epsilon
\ee
where $\m$ runs over all 11 coordinates, and $\hat \g =\g^{0123}$ and 
admits a full compliment of thirty-two independent solutions. 
We denote by $\g_a=e_a^m \G_m$ the tangent space gamma matrices.\\

To proceed, here we will construct the Killing spinors preserved in the $AdS_4
\times \C\cP^3$ background starting from $AdS_4
\times S^7$ background. Some relevant calculations have been performed in
~\cite{Claus:1998,Lu:1998,Skenderis:2002vf}. We however repeat the calculations in our basis and also introduce
projection operators in this basis. We take the following form for the $AdS_4$
\bea
ds_{AdS4}^2 & = & \frac{du^2}{u^2} + u^2 (dx \cdot dx)_3 \\
&=&\frac{du^2}{u^2} + u^2 [ -(dx^0)^2 + (dx^1)^2 + (dx^2)^2 ]
\eea
Calculating the spin-connection from
$\omega^{\hat a \hat b}_{\mu} =e^{\hat a}_{\mu} ( \partial_{\mu} e^{\nu\hat b} + e^{\tau\hat b} \Gamma^{\nu}_{\tau \mu} ) $ we find for the $AdS_4$,
\bea
  &&\omega^{\hat u\hat u}_{u}= 0,\qquad\omega^{\hat u\hat0}_{0}= -u dx_0,\qquad
  \omega^{\hat u\hat1}_{1}=-u dx_1,\\
  &&\omega^{\hat u\hat2}_{2}=-u dx_2,
\eea
The three-dimensional $N=6$ CS theory is conjectured to be dual to 
M-theory on $AdS_4\times S^7/\bZ_k$. To understand the action of the 
$\bZ_k$ orbifold, it is instructive to write the $S^7$ as a circle fibration over 
$\C\cP^3$, where the orbifold acts on the fiber.  
For large $k$ the radius of this ``M-theory circle'' becomes small, so the 
theory can be described in terms of type IIA string theory 
on $AdS_4\times \C\cP^3$ with the metric
\be
ds^2=\frac{R^3}{4k}\left(ds^2_{AdS_4}+4ds^2_{\C\cP^3}\right)\,.
\label{metric}
\ee
The metric on $\C\cP^3$ can be written in terms of four complex projective
coordinates $z_i$ as
\be
ds_{\C\cP^3}^2=\frac{1}{\rho^2}\sum_{i=1}^4 dz_i\,d\bar z_i
-\frac{1}{\rho^4}\bigg|\sum_{i=1}^4 z_i\,d\bar z_i\bigg|^2\,,\qquad
\rho^2=\sum_{i=1}^4|z_i|^2\,.
\ee
A specific representations in terms of angular 
coordinates is obtained by parameterizing $S^7/Z_k$ as~\cite{Cvetic:2000yp,Nishioka:2008gz},
$
z_1=\cos\frac{\xi}{2}\cos\frac{\vartheta_1}{2}\,e^{i(2\varphi_1+\psi+\zeta)/4}\,,
z_2=\cos\frac{\xi}{2}\sin\frac{\vartheta_1}{2}\,e^{i(-2\varphi_1+\psi+\zeta)/4}\,,
z_3=\sin\frac{\xi}{2}\cos\frac{\vartheta_2}{2}\,e^{i(2\varphi_2-\psi+\zeta)/4}\,,
z_4=\sin\frac{\xi}{2}\sin\frac{\vartheta_2}{2}\,e^{i(-2\varphi_2-\psi+\zeta)/4}\,
$
The metric on $S^7$ is then given by
\bea
ds^2_{S^7}&=&\frac{1}{4}\Bigg[
d\xi^2
+\cos^2\frac{\xi}{2}(d\vartheta_1^2+\sin^2\vartheta_1^2\,d\varphi_1^2)
+\sin^2\frac{\xi}{2}(d\vartheta_2^2+\sin^2\vartheta_2^2\,d\varphi_2^2)
\nonumber\\ \hskip1cm
&+&\sin^2\frac{\xi}{2}\cos^2\frac{\xi}{2}
(d\psi+\cos\vartheta_1\,d\varphi_1-\cos\vartheta_2\,d\varphi_2)^2
+\frac{1}{4}(d\zeta+A)^2\, \Bigg]\,,
\label{S7-metric}\\ \nonumber
A&=&\cos\xi\,d\psi+2\cos^2\frac{\alpha}{2}\cos\vartheta_1\,d\varphi_1
+2\sin^2\frac{\xi}{2}\cos\vartheta_2\,d\varphi_2\,.
\label{S7-form}
\eea
The angle $\zeta$ appears only in the last term and if we drop it 
we end up with the metric on $\C\cP^3$,
\bea
ds^2_{\C\cP^3}&=&\frac{1}{4}\bigg[
d\xi^2
+\cos^2\frac{\xi}{2}(d\vartheta_1^2+\sin^2\vartheta_1^2\,d\varphi_1^2)
+\sin^2\frac{\xi}{2}(d\vartheta_2^2+\sin^2\vartheta_2^2\,d\varphi_2^2) \nonumber
\\
&+&\sin^2\frac{\xi}{2}\cos^2\frac{\xi}{2}
(d\chi+\cos\vartheta_1\,d\varphi_1-\cos\vartheta_2\,d\varphi_2)^2\bigg].
\eea
Although, the above metric is slightly different from the one of eqn. (\ref{cp3earlier}) used earlier  in obtaining
solutions, it can be seen later that it does not effect the analysis in this section.

\noindent
The vielbeins coming from $g_{\mu\nu} = e^{\hat a}_{\mu}e^{\hat a}_{\mu}\eta_{{\hat a}{\hat b}}$ for
$AdS_4$ are
\be
  e^{\hat u}_{u}=\frac{R}{2u}\,du,\qquad
  e^{\hat 0}_{0}= \frac{R}{2}u\,dx_0,\qquad e^{\hat 1}_{1}= \frac{Ru}{2}\,dx_1,\qquad
  e^{\hat 2}_{2}=\frac{Ru}{2}\,dx_2,
\ee
and for $S^7$,
\bea
e^4 &=&\frac{R}{2}  d\xi, \qquad 
e^5 =\frac{R}{2}  \cos\frac{\xi}{2}\,d\vartheta_1,\qquad 
e^6 =\frac{R}{2}  \sin\frac{\xi}{2} \, d\vartheta_2, \nonumber \\
e^7 &=& \frac{R}{2}  \cos\frac{\xi}{2}\sin\frac{\xi}{2} \Bigl( \cos \vartheta_1 \,
d\varphi_1 - \cos\vartheta_2\,d\varphi_2 + d\psi \Bigr),\nonumber \\
e^8 &=& \frac{R}{2}  \cos\frac{\xi}{2} \sin\vartheta_1\,d\varphi_1,\quad
e^9 =\frac{R}{2}  \sin\frac{\xi}{2}\sin\vartheta_2 \, d\varphi_2\,,\nonumber\\
e^\natural &=&-\frac{R}{4}\left(d\zeta+2\cos^2\frac{\xi}{2}\cos\vartheta_1\,d\varphi_1
+2\sin^2\frac{\xi}{2}\cos\vartheta_2\,d\varphi_2+\cos\xi\,d\psi\right).
\eea

The full solution for the killing spinor also includes the function
$ h(\C\cP^3)$, which appears from solving the Killing
equation on $ \C\cP^3$. A quicker way to obtain this, is
the equation on the sphere given as
\be
(D_{\alpha} + \half  \Gamma_{\alpha} )\ep = 0,
\ee
where $D_{\alpha}$ is the covariant derivative on $S^7$,
and the solution is given by
\bea
h(\C\cP^3) =  e^{\frac{\xi}{4} ( \hat \g \g_4 - \g_{7\natural}   ) }
e^{\frac{\vartheta_1}{4} (  \hat \g \g_5 - \g_{8\natural}  ) }
e^{\frac{\vartheta_2}{4} ( \g_{79} + \g_{46} ) }
e^{-\frac{\xi_1}{2} \hat \g \g_\natural}
e^{-\frac{\xi_2}{2} \g_{58}}
e^{-\frac{\xi_3}{2} \g_{47}}
e^{-\frac{\xi_4}{2} \g_{69}}
\label{killcp}
\eea
where the $\xi_i$ are given by
$\xi_1=\frac{2\varphi_1+\chi+\zeta}{4}\,,\qquad
\xi_2=\frac{-2\varphi_1+\chi+\zeta}{4}\,,\qquad
\xi_3=\frac{2\varphi_2-\chi+\zeta}{4}\,,\qquad
\xi_4=\frac{-2\varphi_2-\chi+\zeta}{4}\,$.
The Dirac matrices were chosen such that
$\gamma_{012345678 9\natural}=1$. Similar calculations in different coordinate systems were done in 
\cite{Nishioka:2008ib,Drukker:2008zx}. 
To see which Killing spinors survive the orbifolding, one writes the spinor 
$\epsilon_0$ in a basis which diagonalizes
$i\hat\gamma\gamma_\natural\epsilon_0=s_1\epsilon_0\,,\qquad
i\gamma_{58}\epsilon_0=s_2\epsilon_0\,,\qquad
i\gamma_{47}\epsilon_0=s_3\epsilon_0\,,\qquad
i\gamma_{69}\epsilon_0=s_4\epsilon_0\,$. 
All the $s_i$ take values $\pm1$ and by our conventions on the product 
of all the Dirac matrices, the number of negative eigenvalues is even. 
Now consider a shift along the $\zeta$ circle, which changes all the 
angles by $\xi_i\to\xi_i+\delta/4$, the Killing spinors transform 
as
\be
{\mathcal M}\epsilon_0\to {\mathcal M} e^{i\frac{\delta}{8}
(s_1+s_2+s_3+s_4)}\epsilon_0\,.
\ee
This transformation is a symmetry of the Killing spinor when two of the $s_i$ eigenvalues 
are positive and two negative and not when they all have the same sign (unless $\delta$ 
is an integer multiple of $4\pi$). Note that on $S^7$ the radius of the $\zeta$ circle 
is $8\pi$, so the $\bZ_k$ orbifold of $S^7$ is given by taking $\delta=8\pi/k$. 
The allowed values of the $s_i$ are therefore
\be
(s_1,s_2,s_3,s_4)\in\left\{
\begin{matrix}
(+,+,-,-),\ (+,-,+,-),\ (+,-,-,+),\\
(-,+,+,-),\ (-,+,-,+),\ (-,-,+,+)
\end{matrix}
\right\}
\label{signs}
\ee
Each configuration represents four supercharges, so the orbifolding 
breaks $1/4$ of the supercharges (except for $k=1,2$) and leaves 
24 unbroken supersymmetries.\\


\begin{thebibliography}{99}

\bibitem{Maldacena:1997re}
  J.~M.~Maldacena,
  ``The large N limit of superconformal field theories and supergravity,''
  Adv.\ Theor.\ Math.\ Phys.\  {\bf 2}, 231 (1998)
  [Int.\ J.\ Theor.\ Phys.\  {\bf 38}, 1113 (1999)]
  [arXiv:hep-th/9711200].
\bibitem{Witten:1998qj}
  E.~Witten,
  ``Anti-de Sitter space and holography,''
  Adv.\ Theor.\ Math.\ Phys.\  {\bf 2}, 253 (1998)
  [arXiv:hep-th/9802150]
\bibitem{Gubser:1998bc}
  S.~S.~Gubser, I.~R.~Klebanov and A.~M.~Polyakov,
  ``Gauge theory correlators from non-critical string theory,''
  Phys.\ Lett.\  B {\bf 428}, 105 (1998)
  [arXiv:hep-th/9802109].
  


\bibitem{Aharony:2008ug}
  O.~Aharony, O.~Bergman, D.~L.~Jafferis and J.~Maldacena,
  ``$\cN=6$ superconformal Chern-Simons-matter theories, M2-branes and their
  gravity duals,''
  [arXiv:0806.1218[hep-th]].

\bibitem{Schwarz:2004yj}
  J.~H.~Schwarz,
  ``Superconformal Chern-Simons theories,''
  JHEP {\bf 0411}, 078 (2004)
  [arXiv:hep-th/0411077].
 J.~Bagger and N.~Lambert,
  ``Modeling multiple M2's,''
  Phys.\ Rev.\  D {\bf 75}, 045020 (2007)
  [arXiv:hep-th/0611108].
J.~Bagger and N.~Lambert,
  ``Gauge Symmetry and Supersymmetry of Multiple M2-Branes,''
  Phys.\ Rev.\  D {\bf 77}, 065008 (2008)
  [arXiv:0711.0955 [hep-th]].
  A.~Gustavsson,
  ``Algebraic structures on parallel M2-branes,''
  [arXiv:0709.1260 [hep-th]].
  J.~Gomis, D.~Rodriguez-Gomez, M.~Van Raamsdonk and H.~Verlinde,
  ``Supersymmetric Yang-Mills Theory From Lorentzian Three-Algebras,''
  JHEP {\bf 0808}, 094 (2008)
  [arXiv:0806.0738 [hep-th]].
  J.~Bagger and N.~Lambert,
  ``Comments On Multiple M2-branes,''
  JHEP {\bf 0802}, 105 (2008)
  [arXiv:0712.3738 [hep-th]].
  D.~Gaiotto and X.~Yin,
  ``Genus Two Partition Functions of Extremal Conformal Field Theories,''
  JHEP {\bf 0708}, 029 (2007)
  [arXiv:0707.3437 [hep-th]].
  D.~Gaiotto and E.~Witten,
  ``Janus Configurations, Chern-Simons Couplings, And The Theta-Angle in N=4
  [arXiv:0804.2907 [hep-th]].
  M.~A.~Bandres, A.~E.~Lipstein and J.~H.~Schwarz,
  ``N = 8 Superconformal Chern--Simons Theories,''
  JHEP {\bf 0805}, 025 (2008)
  [arXiv:0803.3242 [hep-th]].
  M.~Van Raamsdonk,
  ``Comments on the Bagger-Lambert theory and multiple M2-branes,''
  JHEP {\bf 0805}, 105 (2008)
  [arXiv:0803.3803 [hep-th]].
  N.~Lambert and D.~Tong,
  ``Membranes on an Orbifold,''
  Phys.\ Rev.\ Lett.\  {\bf 101}, 041602 (2008)
  [arXiv:0804.1114 [hep-th]].
  S.~Benvenuti, D.~Rodriguez-Gomez, E.~Tonni and H.~Verlinde,
  ``N=8 superconformal gauge theories and M2 branes,''
  [arXiv:0805.1087 [hep-th]].
  C.~Krishnan and C.~Maccaferri,
  ``Membranes on Calibrations,''
  JHEP {\bf 0807}, 005 (2008)
  [arXiv:0805.3125 [hep-th]].
  M.~A.~Bandres, A.~E.~Lipstein and J.~H.~Schwarz,
  ``Ghost-Free Superconformal Action for Multiple M2-Branes,''
  JHEP {\bf 0807}, 117 (2008)
  [arXiv:0806.0054 [hep-th]].
  J.~Bagger and N.~Lambert,
  ``Three-Algebras and N=6 Chern-Simons Gauge Theories,''
  [arXiv:0807.0163 [hep-th]].


\bibitem{Mukhi:2008ux}
  S.~Mukhi and C.~Papageorgakis,
  ``M2 to D2,''
  JHEP {\bf 0805} (2008) 085
  [arXiv:0803.3218 [hep-th]].

\bibitem{Ezhuthachan:2008ch}
  B.~Ezhuthachan, S.~Mukhi and C.~Papageorgakis,
  ``D2 to D2,''
  JHEP {\bf 0807} (2008) 041
  [arXiv:0806.1639 [hep-th]].





\bibitem{Arutyunov:2008if}
  G.~Arutyunov and S.~Frolov,
  ``Superstrings on $AdS_4 x CP^3$ as a Coset Sigma-model,''
  JHEP {\bf 0809}, 129 (2008)
  [arXiv:0806.4940 [hep-th]].



\bibitem{Stefanski:2008ik}
  B.~.~j.~Stefanski,
  ``Green-Schwarz action for Type IIA strings on $AdS_4\times CP^3$,''
  Nucl.\ Phys.\  B {\bf 808}, 80 (2009)
  [arXiv:0806.4948 [hep-th]].

\bibitem{Fre:2008qc}
  P.~Fre and P.~A.~Grassi,
  ``Pure Spinor Formalism for {Osp}(N|4) backgrounds,''
  arXiv:0807.0044 [hep-th].

\bibitem{Bonelli:2008us}
  G.~Bonelli, P.~A.~Grassi and H.~Safaai,
  ``Exploring Pure Spinor String Theory on $AdS_4\times \mathbb{CP}^3$,''
  JHEP {\bf 0810}, 085 (2008)
  [arXiv:0808.1051 [hep-th]].

\bibitem{D'Auria:2008cw}
  R.~D'Auria, P.~Fre, P.~A.~Grassi and M.~Trigiante,
  ``Superstrings on $AdS_4 x CP^3$ from Supergravity,''
  Phys.\ Rev.\  D {\bf 79}, 086001 (2009)
  [arXiv:0808.1282 [hep-th]].


\bibitem{Giani:1984wc}
  F.~Giani and M.~Pernici,
  ``N=2 Supergravity In Ten-Dimensions,''
  Phys.\ Rev.\  D {\bf 30}, 325 (1984).



\bibitem{Nilsson:1984bj}
  B.~E.~W.~Nilsson and C.~N.~Pope,
  ``Hopf Fibration Of Eleven-Dimensional Supergravity,''
  Class.\ Quant.\ Grav.\  {\bf 1}, 499 (1984).


\bibitem{Sorokin:1985ap1}
D.~P.~Sorokin, V.~I.~Tkach and D.~V.~Volkov, 
``Kaluza-Klein Theories And Spontaneous Compactification Mechanisms Of Extra Space
Dimensions,'' {\it  In *Moscow 1984, Proceedings, Quantum Gravity*, 376-392}


\bibitem{Sorokin:1985ap}
  D.~P.~Sorokin, V.~I.~Tkach and D.~V.~Volkov,
  ``On The Relationship Between Compactified Vacua Of D = 11 And D = 10
  Supergravities,''
  Phys.\ Lett.\  B {\bf 161}, 301 (1985).


\bibitem{Gomis:2008jt}
  J.~Gomis, D.~Sorokin and L.~Wulff,
  ``The complete $AdS(4) x CP(3)$ superspace for the type IIA superstring and
  D-branes,''
  JHEP {\bf 0903}, 015 (2009)
  [arXiv:0811.1566 [hep-th]].

\bibitem{Grassi:2009yj}
  P.~A.~Grassi, D.~Sorokin and L.~Wulff,
  ``Simplifying superstring and D-brane actions in $AdS(4) x CP(3)$
  superbackground,''
  JHEP {\bf 0908}, 060 (2009)
  [arXiv:0903.5407 [hep-th]].

\bibitem{Cagnazzo:2009zh}
  A.~Cagnazzo, D.~Sorokin and L.~Wulff,
  ``String instanton in AdS(4)xCP(3),''
  JHEP {\bf 1005}, 009 (2010)
  [arXiv:0911.5228 [hep-th]].


\bibitem{Mirabelli:1997aj}
  E.~A.~Mirabelli and M.~E.~Peskin,
  ``Transmission of supersymmetry breaking from a 4-dimensional boundary,''
  Phys.\ Rev.\  D {\bf 58}, 065002 (1998)
  [arXiv:hep-th/9712214].

  
\bibitem{bachas2}
C.~Bachas and M.~Petropoulos, ``Anti-de Sitter D-branes,'' JHEP
{\bf 0102}, 025 (2001) [arXiv:hep-th/0012234].

\bibitem{randall}
A.~Karch and L.~Randall, ``Locally localized gravity,'' JHEP {\bf 0105},
  008 (2001) [arXiv: hep-th/0011156].
\bibitem{KR} 
A.~Karch and L.~Randall,
``Open and closed string interpretation of SUSY CFT's on branes 
with  boundaries,''
JHEP {\bf 0106}, 063 (2001)
[arXiv:hep-th/0105132].

\bibitem{Randall:1999vf}
  L.~Randall and R.~Sundrum,
  ``An alternative to compactification,''
  Phys.\ Rev.\ Lett.\  {\bf 83}, 4690 (1999)
  [arXiv:hep-th/9906064].




\bibitem{ooguri1}
O.~DeWolfe, D.~Freedman, and H.~Ooguri, ``Holography and defect conformal
  field theories,''[ arXiv:  hep-th/0111135].

\bibitem{bachas1}
C.~Bachas, J.~de Boer, R.~Dijkgraaf and H.~Ooguri, ``Permeable
conformal walls and holography,'' [arXiv:hep-th/0111210].


\bibitem{lee}
P.~Lee, J.~Park, and H.~Ooguri, ``Boundary states for $AdS_2$ branes in
  $AdS_3$,''[ arXiv: hep-th/0112188].

\bibitem{ponsot}
B.~Ponsot, V.~Schomerus, and J.~Teschner, ``Branes in the Euclidean $AdS_3$,''
  {\em JHEP} {\bf 0202}, 016 (2002) [arXiv:hep-th/0112198].

\bibitem{johanna}
J.~Erdmenger, Z.~Guralnik and I.~Kirsch,
``Four-dimensional superconformal theories with interacting 
boundaries or  defects,''
[arXiv:hep-th/0203020].

\bibitem{Karch:2002sh}
  A.~Karch and E.~Katz,
  ``Adding flavor to AdS/CFT,''
  JHEP {\bf 0206}, 043 (2002)
  [arXiv:hep-th/0205236].

\bibitem{Myers:2006qr}
  R.~C.~Myers and R.~M.~Thomson,
  ``Holographic mesons in various dimensions,''
  JHEP {\bf 0609}, 066 (2006)
  [arXiv:hep-th/0605017].



\bibitem{Fujita:2009kw}
  M.~Fujita, W.~Li, S.~Ryu and T.~Takayanagi,
  ``Fractional Quantum Hall Effect via Holography: Chern-Simons, Edge States,
  and Hierarchy,''
  JHEP {\bf 0906}, 066 (2009)
  [arXiv:0901.0924 [hep-th]].


\bibitem{Hikida:2009tp}
  Y.~Hikida, W.~Li and T.~Takayanagi,
  ``ABJM with Flavors and FQHE,''
  JHEP {\bf 0907}, 065 (2009)
  [arXiv:0903.2194 [hep-th]].



\bibitem{Hohenegger:2009as}
  S.~Hohenegger and I.~Kirsch,
  ``A note on the holography of Chern-Simons matter theories with flavour,''
  JHEP {\bf 0904}, 129 (2009)
  [arXiv:0903.1730 [hep-th]].
  
\bibitem{Gaiotto:2009mv}
  D.~Gaiotto and A.~Tomasiello,
  ``The gauge dual of Romans mass,''
  arXiv:0901.0969 [hep-th].
 

\bibitem{Gaiotto:2009tk}
  D.~Gaiotto and D.~L.~Jafferis,
  ``Notes on adding D6 branes wrapping RP3 in $AdS4 x CP3$,''
  arXiv:0903.2175 [hep-th].
  
\bibitem{Ammon:2009wc}
  M.~Ammon, J.~Erdmenger, R.~Meyer, A.~O'Bannon and T.~Wrase,
  ``Adding Flavor to AdS4/CFT3,''
  arXiv:0909.3845 [hep-th].
  
\bibitem{Benini:2009qs}
  F.~Benini, C.~Closset and S.~Cremonesi,
  ``Chiral flavors and M2-branes at toric CY4 singularities,''
  JHEP {\bf 1002}, 036 (2010)
  [arXiv:0911.4127 [hep-th]].

  
\bibitem{Jafferis:2009th}
  D.~L.~Jafferis,
  ``Quantum corrections to N=2 Chern-Simons theories with flavor and their AdS4
  duals,''
  arXiv:0911.4324 [hep-th].
  

\bibitem{Fujita:2010pj}
  M.~Fujita,
  ``M5-brane Defect and QHE in $AdS_4 \times N(1,1)/N=3$ SCFT,''
  arXiv:1011.0154 [hep-th].


\bibitem{Koerber:2009he}
  P.~Koerber,
  ``Coisotropic D-branes on $AdS4 x CP3$ and massive deformations,''
  JHEP {\bf 0909}, 008 (2009)
  [arXiv:0904.0012 [hep-th]].

\bibitem{Koerber:2007jb}
  P.~Koerber and L.~Martucci,
  ``D-branes on AdS flux compactifications,''
  JHEP {\bf 0801}, 047 (2008)
  [arXiv:0710.5530 [hep-th]].

\bibitem{Chen:2008bp}
  B.~Chen and J.~B.~Wu,
  ``Supersymmetric Wilson Loops in N=6 Super Chern-Simons-matter theory,''
  Nucl.\ Phys.\  B {\bf 825}, 38 (2010)
  [arXiv:0809.2863 [hep-th]].
  S.~J.~Rey, T.~Suyama and S.~Yamaguchi,
  ``Wilson Loops in Superconformal Chern-Simons Theory and Fundamental Strings
  in Anti-de Sitter Supergravity Dual,''
  JHEP {\bf 0903}, 127 (2009)
  [arXiv:0809.3786 [hep-th]].
  N.~Drukker, J.~Gomis and D.~Young,
  ``Vortex Loop Operators, M2-branes and Holography,''
  JHEP {\bf 0903}, 004 (2009)
  [arXiv:0810.4344 [hep-th]].
  J.~Kluson and K.~L.~Panigrahi,
  ``Defects and Wilson Loops in 3d QFT from D-branes in $AdS(4) x CP**3$,''
  arXiv:0809.3355 [hep-th].
  
\bibitem{Skenderis:2002vf}
  K.~Skenderis and M.~Taylor,
  ``Branes in AdS and pp-wave spacetimes,''
  JHEP {\bf 0206} (2002) 025
  [arXiv:hep-th/0204054].
  
  
\bibitem{Skenderis:2002wx}
  K.~Skenderis and M.~Taylor,
  ``Open strings in the plane wave background. I: Quantization and
  symmetries,''
  Nucl.\ Phys.\  B {\bf 665}, 3 (2003)
  [arXiv:hep-th/0211011].
  
\bibitem{Skenderis:2002ps}
  K.~Skenderis and M.~Taylor,
  ``Open strings in the plane wave background. II: Superalgebras and spectra,''
  JHEP {\bf 0307}, 006 (2003)
  [arXiv:hep-th/0212184].

 
  
\bibitem{AliAkbari:2010rs}
  M.~Ali-Akbari,
  ``A D2-brane in the Penrose limits of AdS(4)x CP(3),''
  Phys.\ Rev.\  D {\bf 82}, 065027 (2010)
  [arXiv:1005.0126 [hep-th]].
  
\bibitem{Drukker:2008zx}
  N.~Drukker, J.~Plefka and D.~Young,
  ``Wilson loops in 3-dimensional N=6 supersymmetric Chern-Simons Theory and
  their string theory duals,''
  JHEP {\bf 0811}, 019 (2008)
  [arXiv:0809.2787 [hep-th]].
  

  
\bibitem{Nishioka:2008gz}
  T.~Nishioka and T.~Takayanagi,
  ``On Type IIA Penrose Limit and $\cN=6$ Chern-Simons Theories,''
  JHEP {\bf 0808} (2008) 001
  [arXiv:0806.3391[hep-th]].


\bibitem{Karch:2000gx}
  A.~Karch and L.~Randall,
  ``Open and closed string interpretation of SUSY CFT's on branes with
  boundaries,''
  JHEP {\bf 0106}, 063 (2001)
  [arXiv:hep-th/0105132].


\bibitem{Breitenlohner:1982bm}
  P.~Breitenlohner and D.~Z.~Freedman,
  ``Positive Energy In Anti-De Sitter Backgrounds And Gauged Extended
  Supergravity,''
  Phys.\ Lett.\  B {\bf 115}, 197 (1982).
  
\bibitem{Bachas:1999um}
C.~P.~Bachas, P.~Bain and M.~B.~Green,
``Curvature terms in D-brane actions and their M-theory origin,''
JHEP {\bf 9905}, 011 (1999)
[arXiv:hep-th/9903210].
  
  
\bibitem{Harvey:2007ab}
  J.~A.~Harvey and A.~B.~Royston,
  ``Localized Modes at a D-brane--O-plane Intersection and Heterotic Alice
  Strings,''
  JHEP {\bf 0804}, 018 (2008)
  [arXiv:0709.1482 [hep-th]].

\bibitem{Buchbinder:2007ar}
  E.~I.~Buchbinder, J.~Gomis and F.~Passerini,
  ``Holographic Gauge Theories in Background Fields and Surface Operators,''
  JHEP {\bf 0712}, 101 (2007)
  [arXiv:0710.5170 [hep-th]].
  
\bibitem{Harvey:2008zz}
  J.~A.~Harvey and A.~B.~Royston,
  ``Gauge/Gravity duality with a chiral N=(0,8) string defect,''
  JHEP {\bf 0808}, 006 (2008)
  [arXiv:0804.2854 [hep-th]].

\bibitem{Aharony:2008gk}
  O.~Aharony, O.~Bergman and D.~L.~Jafferis,
  ``Fractional M2-branes,''
  JHEP {\bf 0811}, 043 (2008)
  [arXiv:0807.4924 [hep-th]].
  
\bibitem{Jensen:2010vx}
  K.~Jensen,
  ``More Holographic Berezinskii-Kosterlitz-Thouless Transitions,''
  Phys.\ Rev.\  D {\bf 82}, 046005 (2010)
  [arXiv:1006.3066 [hep-th]].
  
\bibitem{Fiol:2010un}
  B.~Fiol,
  ``Flavor from M5-branes,''
  JHEP {\bf 1007}, 046 (2010)
  [arXiv:1005.2133 [hep-th]].



 
\bibitem{Lee:2002cu}
  P.~Lee and J.~w.~Park,
  ``Open strings in PP-wave background from defect conformal field theory,''
  Phys.\ Rev.\  D {\bf 67}, 026002 (2003)
  [arXiv:hep-th/0203257].
  
\bibitem{Yamaguchi:2003ay}
  S.~Yamaguchi,
  ``AdS branes corresponding to superconformal defects,''
  JHEP {\bf 0306}, 002 (2003)
  [arXiv:hep-th/0305007].
  
\bibitem{Lunin:2007ab}
  O.~Lunin,
  ``1/2-BPS states in M theory and defects in the dual CFTs,''
  JHEP {\bf 0710}, 014 (2007)
  [arXiv:0704.3442 [hep-th]].
  
  
  
 
\bibitem{Sethi:1997zz}
S.~Sethi,
``The matrix formulation of type IIB five-branes,''
Nucl.\ Phys.\ B {\bf 523} (1998) 158
[arXiv:hep-th/9710005].

\bibitem{Ganor:1997jx}
O.~J.~Ganor and S.~Sethi,
``New perspectives on Yang-Mills theories with sixteen supersymmetries,''
JHEP {\bf 9801} (1998) 007
[arXiv:hep-th/9712071].

\bibitem{Kapustin:1998pb}
A.~Kapustin and S.~Sethi,
Adv.\ Theor.\ Math.\ Phys.\  {\bf 2} (1998) 571
[arXiv:hep-th/9804027].

\bibitem{Hori:2000ic}
K.~Hori,
``Linear models of supersymmetric D-branes,''
arXiv:hep-th/0012179.




  
  
\bibitem{Hartnoll:2007ai}
  S.~A.~Hartnoll and P.~Kovtun,
  ``Hall conductivity from dyonic black holes,''
  Phys.\ Rev.\  D {\bf 76}, 066001 (2007)
  [arXiv:0704.1160 [hep-th]].
  
\bibitem{Jensen:2010ga}
  K.~Jensen, A.~Karch, D.~T.~Son and E.~G.~Thompson,
  ``Holographic Berezinskii-Kosterlitz-Thouless Transitions,''
  Phys.\ Rev.\ Lett.\  {\bf 105}, 041601 (2010)
  [arXiv:1002.3159 [hep-th]].
  

 
\bibitem{Breitenlohner:1982jf}
  P.~Breitenlohner and D.~Z.~Freedman,
  ``Stability In Gauged Extended Supergravity,''
  Annals Phys.\  {\bf 144}, 249 (1982).


  

\bibitem{Evans:2010hi}
  N.~Evans, A.~Gebauer, K.~Y.~Kim and M.~Magou,
  ``Phase diagram of the D3/D5 system in a magnetic field and a BKT
  transition,''
  arXiv:1003.2694 [hep-th].








\bibitem{Cederwall:1996pv}
M.~Cederwall, A.~von Gussich, B.~Nilsson and A.~Westerberg,
``The Dirichlet super-three-brane in ten-dimensional type IIB supergravity,''
Nucl.\ Phys.\ B {\bf 490}, 163 (1997)
[arXiv:hep-th/9610148].


\bibitem{Aganagic:1996pe}
M.~Aganagic, C.~Popescu and J.~Schwarz,
``D-brane actions with local kappa symmetry,''
Phys.\ Lett.\ B {\bf 393}, 311 (1997)
[arXiv:hep-th/9610249].

\bibitem{Cederwall:1996ri}
M.~Cederwall, A.~von Gussich, B.~Nilsson, P.~Sundell and A.~Westerberg,
``The Dirichlet super-p-branes in ten-dimensional type IIA and IIB  
supergravity,''
Nucl.\ Phys.\ B {\bf 490}, 179 (1997)
[arXiv:hep-th/9611159].

\bibitem{Bergshoeff:1996tu}
E.~Bergshoeff and P.~Townsend, ``Super D-branes,''
Nucl.\ Phys.\ B {\bf 490}, 145 (1997)
[arXiv:hep-th/9611173].

\bibitem{Aganagic:1996nn}
M.~Aganagic, C.~Popescu and J.~Schwarz,
``Gauge-invariant and gauge-fixed D-brane actions,''
Nucl.\ Phys.\ B {\bf 495}, 99 (1997)
[arXiv:hep-th/9612080].

\bibitem{Bergshoeff:1997kr}
E.~Bergshoeff, R.~Kallosh, T.~Ortin and G.~Papadopoulos,
``Kappa-symmetry, supersymmetry and intersecting branes,''
Nucl.\ Phys.\ B {\bf 502}, 149 (1997)
[arXiv:hep-th/9705040].

\bibitem{Claus:1998}
P.~Claus and R.~Kallosh, 
``Superisometries of the AdS*S superspace,''
JHEP {\bf 9903}, 014 (1999)
[arXiv:hep-th/9812087].

\bibitem{Lu:1998}
H.~L\"{u}, C.~Pope and J.~Rahmfeld,
``A construction of Killing spinors on $S^n$,''
J.\ Math.\ Phys. {\bf 40}, 4518 (1999)
[arXiv:hep-th/9805151].

  
\bibitem{Cvetic:2000yp}
  M.~Cvetic, H.~Lu and C.~N.~Pope,
  ``Consistent warped-space Kaluza-Klein reductions, half-maximal gauged
  supergravities and CP(n) constructions,''
  Nucl.\ Phys.\  B {\bf 597} (2001) 172
  [hep-th/0007109]. 
  S.~Watamura,
  ``Spontaneous Compactification And Cp(N): SU(3) X SU(2) X U(1),
  Sin**2-Theta-W, G(3) / G(2) And SU(3) Triplet Chiral Fermions In
  Four-Dimensions,''
  Phys.\ Lett.\  B {\bf 136}, 245 (1984).

 


\bibitem{Nishioka:2008ib}
  T.~Nishioka and T.~Takayanagi,
  ``Fuzzy Ring from M2-brane Giant Torus,''
  [arXiv:0808.2691[hep-th]].













\end{thebibliography}
\end{document}